\documentclass[12pt]{iopart}

\usepackage{citesort,latexsym,ifthen,graphics,color}

\newcommand{\ie}{\emph{i.e.}}
\newcommand{\cf}{\emph{cf.}}
\newcommand{\eg}{\emph{e.g.}}
\newcommand{\aka}{\emph{a.k.a.}}

\newcommand{\viz}{\emph{viz.}}
\newcommand{\vev}{\emph{vev}}

\newcommand{\wrt}{with respect to}

\newcommand{\lhs}{left-hand side}
\newcommand{\rhs}{right-hand side}

\newcommand{\be}{\begin{equation}}
\newcommand{\ee}{\end{equation}}
\newcommand{\bea}{\begin{eqnarray}}
\newcommand{\eea}{\end{eqnarray}}
\newcommand{\beas}{\begin{eqnarray*}}
\newcommand{\eeas}{\end{eqnarray*}}

\newcommand{\bear}{\begin{array}{l}}
\newcommand{\eear}{\end{array}}

\newcommand{\bcf}{\begin{center}\begin{figure}}
\newcommand{\ecf}{\end{figure}\end{center}}

\newcommand{\bct}{\begin{center}\begin{table}}
\newcommand{\ect}{\end{table}\end{center}}

\newcommand{\ds}{\displaystyle}

\def\A{{\cal A}}
\def\C{{\cal C}}

\def\F{{\cal F}}

\def\P{{\cal P}}

\def\eq#1{(\ref{eq:#1})}

\def\Eqn#1{Equation~(\ref{eq:#1})}

\def\eqs#1#2{(\ref{eq:#1}) and~(\ref{eq:#2})}

\def\sec#1{section~\ref{sec:#1}}

\def\fig#1{figure~\ref{fig:#1}}

\newcommand{\inte}{\! \int \!\!}
\newcommand{\nCr}[2]{ \left.\right.^{#1}\!C_{#2}\ }
\newcommand{\Delslash}{\! \not{\hspace{-0.2em} \nabla}}

\newcommand{\Int}[1]{\int \!\! d^D \! #1 \,}
\newcommand{\volume}[1]{d^D \! #1 \,}

\newcommand{\pder}[2]{\ensuremath{\frac{\partial #1}{\partial #2}}}
\newcommand{\fder}[2]{\ensuremath{\frac{\delta #1}{\delta #2}}}
\newcommand{\Op}[1]{\Or (p^{#1})}

\newcommand{\hf}{\frac{1}{2}}

\newcommand{\expectation}[1]{\left\langle #1 \right\rangle}
\newcommand{\measure}[1]{\mathcal{D} #1 \, }

\newcommand{\bindexed}[3]{\left( #1 \right)^{#2}_{#3}}

\newcommand{\SU}{\mathrm{SU}}
\newcommand{\U}{\mathrm{U}}
\newcommand{\up}{\psi_u}
\newcommand{\down}{\psi_d}
\newcommand{\antiup}{\bar{\psi}_u}
\newcommand{\antidown}{\bar{\psi}_d}
\newcommand{\Up}{\Psi_u}
\newcommand{\Down}{\Psi_d}
\newcommand{\antiUp}{\bar{\Psi}_u}
\newcommand{\antiDown}{\bar{\Psi}_d}
\newcommand{\Quarks}{\Psi}
\newcommand{\quarks}{\psi}

\def\dd{\dot{\Delta}}
\def\ker#1{\!\cdot\! #1 \!\cdot\!}
\def\hS{\hat{S}}
\def\e#1{\,{\rm e}^{\displaystyle #1}}
\def\one{\hbox{1\kern-.8mm l}}
\def\str{\mathrm{str}\,}

\def\diag{\mathrm{diag}\,}

\newcommand{\CovKer}[2]{\{#1\}_{\!\!{}_{#2}}}
\newcommand{\TiedCovKer}[4]{#1 \, \CovKer{#2}{#3} #4}
\newcommand{\GR}{\cdeps{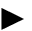}}
\newcommand{\GRk}{\rhd}
\newcommand{\GRkpr}{>}
\newcommand{\DiagDot}{\scriptstyle \bullet}
\newcommand{\DummyKernel}{\ensuremath{\stackrel{\bullet}{\mbox{\rule{1cm}{.2mm}}}}}

\newcommand{\flow}{\Lambda \partial_\Lambda}

\newcommand{\flowConstAlj}{\Lambda \partial_\Lambda|_{\overline{m}_u, \overline{m}_c, \ldots}}
\newcommand{\dec}[3][0]{\ensuremath{\left[ #2 \hspace{#1em} \right]^{#3}}}

\newlength{\PFheight}
\newcommand{\PushF}[1]{
	 \stackrel{\rightarrow}{\vspace{\PFheight} #1}
}

\newcommand{\PullB}[1]{
	\stackrel{\leftarrow}{\vspace{\PFheight} #1}
}

\newcommand{\cdeps}[1]{\ensuremath{\begin{array}{c}\includegraphics{./eps/#1.eps} \end{array}}}

\newcommand{\jhep}[3]{{JHEP} #1 (#2) #3}
\newcommand{\NuclPhys}[4]{{Nucl.\ Phys.\ }\textbf{#1 #2} (#3) #4}
\newcommand{\PhysRev}[4]{{Phys.\ Rev.\ }\textbf{#1 #2} (#3) #4}
\newcommand{\IntJModPhys}[4]{{Int.\ J.\ Mod.\ Phys.\ }\textbf{#1 #2} (#3) #4}
\newcommand{\PhysRep}[4]{{Phys.\ Rep.\ }\textbf{#1 #2} (#3) #4}
\newcommand{\PhysRept}[3]{{Phys.\ Rept.\ }\textbf{#1} (#2) #3}
\newcommand{\TheorMathPhys}[3]{{Theor.\ Math.\ Phys.\ }\textbf{#1} (#2) #3}

\newcommand{\PhysLett}[4]{{Phys.\ Lett.\ }\textbf{#1 #2} (#3) #4}
\newcommand{\ProgTheorPhys}[3]{{Prog.\ Theor.\ Phys.\ }\textbf{#1} (#2) #3}
\newcommand{\ProgTheorPhysS}[3]{{Prog.\ Theor.\ Phys.\ Suppl.\ }\textbf{#1} (#2) #3}

\newcommand{\CEurJPhys}[3]{{Central Eur.\ J.\ Phys.\ }\textbf{#1} (#2) #3}

\newcommand{\arxiv}[1]{#1}
\newcommand{\hepth}[1]{hep-th/#1}
\newcommand{\hepph}[1]{hep-ph/#1}

\newcommand{\condmat}[1]{cond-mat/#1}

\newcommand{\RevModPhys}[3]{{Rev.\ Mod.\ Phys.\ }\textbf{#1} (#2) #3}

\newcommand{\jphysa}[3]{J.\ Phys.\ {\bf A}: Math.\ Gen.\ #1 (#2) #3}

\eqnobysec

\begin{document}

\title{Manifestly Gauge Invariant QCD}

\author{
	Tim R.~Morris, Oliver J.~Rosten
}

\address{School of Physics and Astronomy,  University of Southampton,
	Highfield, Southampton SO17 1BJ, U.K.}
\eads{\mailto{T.R.Morris@soton.ac.uk}, \mailto{O.J.Rosten@soton.ac.uk}}

\begin{abstract}
	Building on recent work in $\SU(N)$ Yang-Mills theory,
	we construct a manifestly gauge invariant exact
	renormalization group for QCD. A gauge invariant
	cutoff is constructed by embedding the physical
	gauge theory in a spontaneously broken $\SU(N|N)$
	gauge theory, regularized by covariant higher
	derivatives. Intriguingly, the
	construction is most efficient if the number of
	flavours is a multiple of the number of colours.
	The formalism is illustrated with a very compact 
	calculation of the one-loop $\beta$ function,
	achieving a manifestly universal result and
	without fixing the gauge.
\end{abstract}

\vspace{-60ex}
\hfill SHEP 06-20
\vspace{61ex}

\pacs{11.15.Tk, 11.10.Hi, 12.38.-t}
\maketitle

\setcounter{tocdepth}{2}
\tableofcontents
\markboth{Manifestly Gauge Invariant QCD}{Manifestly Gauge Invariant QCD}

\section{Introduction}

One of the hallmarks of QCD is the qualitatively very different
behaviours observed in the high and low energy domains. In
the former case, the theory exhibits asymptotic freedom, allowing
phenomena to be described in terms 
of weakly interacting quarks and gluons. Since the coupling is
small, calculations can be performed in perturbation theory.
However, as the energy scale is lowered, so the 
coupling strength increases, ultimately
causing the quarks and gluons to be bound together into
hadrons. The failure of perturbative techniques
to capture this behaviour presents a stern challenge.

A promising approach to extracting information from the strongly
coupled (or non-perturbative) domain of quantum field theories
is the exact renormalization
group (ERG)~\cite{Wil,W&H,Pol}, the continuous version of Wilson's
RG. The central feature of the ERG is the implementation
of a momentum cutoff, $\Lambda$, in the theory  
in such a way that the physics
at this scale---which is encoded in the Wilsonian effective
action, $S_\Lambda$---is described in terms of parameters relevant
to this scale. The ERG equation determines how $S_\Lambda$ evolves
with $\Lambda$, thereby linking physics at different
energy scales. Consequently, an ERG for QCD has the potential to
provide access to the strongly coupled regime.

In addition to providing a powerful framework for addressing
a wealth of non-perturbative problems in a range of settings
(see~\cite{Fisher:1998kv,TRM-elements,Litim:1998nf,Aoki:2000wm,Berges:2000ew,Bagnuls:2000ae,Polonyi:2001se,Salmhofer:2001tr,Delamotte:2003dw} 
for reviews), a particular advantage conferred by the ERG
is the huge freedom inherent in its construction~\cite{jose}.
In the context of gauge theories, this freedom can
be exploited to construct manifestly gauge invariant 
ERGs~\cite{ym,ymi,ymii,aprop,Thesis,mgierg1} (for a comprehensive review of the alternatives,
see~\cite{Pawlowski:2005xe}).
Whilst being of obvious novelty value, manifest gauge
invariance also provides both technical and conceptual benefits.
From the technical standpoint,
the gauge field is protected from field strength renormalization and
the Ward identities take a particularly simple form
since the Wilsonian effective action is built only from
gauge invariant combinations of the covariant derivative,
even at the quantum level~\cite{ymi}. In addition, the difficult
technical issue of
Gribov copies~\cite{Gribov} is entirely avoided.
Conceptually, a strong case can be made for manifest
gauge invariance being the natural language to describe
non-perturbative phenomena, not least because 
all conclusions drawn
will be completely gauge independent.

The majority of work into this scheme has, so far, focused on
constructing~\cite{ym,ymi,ymii,aprop,Thesis,mgierg1},
testing~\cite{ymii,aprop,Thesis,mgierg2} and refining~\cite{Primer,RG2005,mgiuc} 
the formalism. Most recently, however, real progress
has been made in understanding how to compute objects
of particular interest; specifically, the expectation values of
gauge invariant operators~\cite{evalues}. 
Moreover, crucial steps in this procedure
have a non-perturbative extension~\cite{InPrep}.
Given
these developments we feel that it is timely to
extend the framework to incorporate quarks, in anticipation
of application to non-perturbative QCD.

The strategy we adopt is to build directly on to the $\SU(N)$ Yang-Mills
construction, which we briefly describe. 
Recall that the implementation of a gauge
invariant cutoff comprises two ingredients~\cite{SU(N|N)}. First, we
apply covariant higher derivative regularization. However,
as is well known~\cite{One-loop-UVDivs}, this is insufficient to completely
regularize the theory, since certain one-loop divergences
slip through. The solution we employ is to instead apply
the covariant higher derivative regularization to a spontaneously
broken $\SU(N|N)$ gauge theory, into which the physical $\SU(N)$ gauge
theory has been embedded. The
heavy fields arising from the symmetry breaking act as
Pauli-Villars (PV) fields, supplementing the covariant
higher derivatives to furnish a complete regularization
of the physical theory.

The symmetry breaking is carried by a Higgs field, $\C$.
Upon acquiring a vacuum expectation value (\vev), 
this field breaks $\SU(N|N)$ down to its bosonic
subgroup, $\SU(N)\times \SU(N)\times \U(1)$. One of these
$\SU(N)$ symmetries is identified with the symmetry of the
physical gauge field, $A^1$; the other
is identified with an unphysical field, $A^2$,
which we note comes with wrong sign action. (We can
effectively ignore the $\U(1)$, as we will see later.) 
Besides $A^1$, only $A^2$ remains massless upon 
spontaneous symmetry breaking, all other
fields picking up a mass of order $\Lambda$. As we send $\Lambda$
to infinity, all effective interactions between
$A^1$ and $A^2$ vanish and so the non-unitary $A^2$
sector decouples and can be ignored~\cite{SU(N|N)}.

The most obvious way to add quarks is to embed them in the
fundamental representation of $\SU(N|N)$. However,
as we will see, the structure of the $\SU(N|N)$ group in fact
forbids this program. 
Instead, we embed $N$ quarks into a field
which transforms as a bifundamental under $\U(N|N)$.
This essentially corresponds to a
gauging of not only the  physical colour symmetry 
but also, in  an entirely unphysical way, a
flavour symmetry. 
The unphysical fields which accompany
the physical quarks are given a mass of order
the cutoff and so act as precisely the set of PV fields
we require for the theory to be regularized.
In this way, we are able to incorporate quarks 
in multiples of $N$, such that the elements of
each set have the same mass. 
To obtain quarks with arbitrary masses, we now 
modify the Higgs sector to 
break the unphysical $\SU(N)$ symmetry
completely. 

Quite apart from the ERG, the inclusion of quarks in
the $\SU(N|N)$ regularizing scheme is of interest in 
itself, since it allows the construction of a real,
gauge invariant cutoff in QCD. In the case of pure
$\mathcal{N}= 4$ super Yang-Mills, a dual picture
of this was recently constructed~\cite{EMR}, 
providing a concrete understanding of how the radial
direction in the AdS/CFT correspondence plays the
role of a gauge invariant measure of energy scale.
It will be interesting to see whether the inclusion
of quarks in the $\SU(N|N)$ regularizing scheme leads
to a new way to introduce quarks in the dual picture.

Having incorporated massive quarks into the regularization
framework, it is now a straightforward matter to 
generalize our manifestly gauge invariant flow equation
from $\SU(N)$ Yang-Mills to QCD. To understand how we go about
doing this, we review the structure of general flow equations.
One of the key ingredients of any
flow equation is that the partition function
(and hence the physics derived from it)
is invariant under the flow. As a consequence of this, the family of
flow equations for some generic fields, $\varphi$, follows
from~\cite{ym,ymi,mgierg1,jose}
\be
\label{eq:blocked}
-\flow \e{-S[\varphi]} =  \int_x \fder{}{\varphi(x)} \left(\Psi_x[\varphi] \e{-S[\varphi]}\right),
\ee
where the functional $\Psi$ parameterizes how the high
energy modes are averaged over (and we have written
$S_\Lambda$ as just $S$). The total derivative on
the \rhs\ ensures that the 
partition function $Z = \int \measure{\varphi} \e{-S}$
is invariant under the flow.

Taking $\varphi$ to represent a single scalar field,
we can use~\cite{scalar2}
\be
\label{eq:Psi_x}
	\Psi_x = \hf \int_y \dd_{xy} \fder{\Sigma_1}{\varphi(y)},
\ee
where $\dd$  is an ERG kernel and $\Sigma_1 = S - 2\hat{S}$,
$\hS$ being the seed action~\cite{Thesis,mgierg1,mgierg2,scalar1,scalar2,aprop,giqed}:
a functional which respects the same symmetries as the Wilsonian effective 
action, $S$, and has the same structure. 
Physically, the seed action can be thought of as (partially)
parameterizing a general Kadanoff blocking in the 
continuum~\cite{jose,mgierg1}, and acts as an input
to our flow equation.

There is a deep relationship between the kernel
and the classical, two-point vertices. 
Specifically,
if we set 
the seed action classical, two-point
vertex
equal to its Wilsonian effective action 
counterpart,  $S_0^{\ \varphi\varphi}$, 
then we find that~\cite{scalar2}
\be
\label{eq:scalar-EP}
	S_0^{\ \varphi\varphi} \Delta = 1,
\ee
where $\Delta$ is the integrated kernel \aka\ effective propagator:
\[
	\dd \equiv -\flow \Delta.
\]
\Eqn{scalar-EP} is the effective propagator relation~\cite{aprop}
and is at the heart of the computational technique employed
within our 
approach~\cite{aprop,scalar2,mgierg1,mgierg2,Thesis,Primer,RG2005,mgiuc,evalues,giqed}.

Now suppose that we consider a flow equation for some set of
fields, rather than a single field. It is highly desirable to
insist on an effective propagator relation for each individual
field which means that, in general, the number of effective
propagators---and hence the number of kernels---must equal
the number fields. This observation holds
the key to generalizing our flow equation for $\SU(N)$ Yang-Mills
to one appropriate for QCD.

To construct an ERG for $\SU(N)$ Yang-Mills, we use
the template~\eq{blocked}, covariantize the 
relationship~\eq{Psi_x} and incorporate the $\SU(N|N)$
regularizing structure~\cite{aprop,mgierg1}. 
By defining the covariantization
appropriately (we will review this in \sec{flowreview}), 
we can ensure that
an effective propagator exists for each of the broken
phase fields. Note, however, that the form of the effective
propagator relation is different to~\eq{scalar-EP}
in the gauge sector, as a consequence of the manifest
gauge invariance:
\be
\label{eq:A1-GR}
	S^{\ A^1 A^1}_{0 \mu \ \; \alpha} (p) \Delta^{A^1 A^1}_{\alpha \ \, \nu} (p) = \delta_{\mu\nu} - \frac{p_\mu p_\nu}{p^2}.
\ee
$S^{\ A^1 A^1}_{0 \mu \ \; \alpha} (p)$ is the two-point classical
vertex in the $A^1$ sector, carrying momentum $p$, and 
$\Delta^{A^1 A^1}_{\alpha \ \, \nu} (p)$ is the associated effective
propagator. Thus we see that the effective propagator is
the inverse of the classical, two-point vertex only in the
transverse space; equivalently, it is the inverse only
up to a remainder which we call a `gauge remainder'.

To add quarks, we again use the template~\eq{blocked},
covariantize the 
relationship~\eq{Psi_x} and incorporate the $\SU(N|N)$
regularizing structure, but this time the covariantization
is designed appropriately for the quarks. To allow independent
quark masses, we update the Higgs sector of the
flow  equation and modify
the covariantizations in the gauge and quark sectors
to ensure that there are enough independent kernels
to cope with the breaking of the unphysical $\SU(N)$
symmetry.

Anticipating that the  flow equation for 
QCD is most efficiently stated via 
its diagrammatic 
representation~\cite{aprop,mgierg1,mgierg2,Thesis,Primer,RG2005,mgiuc}, 
we could jump straight to this diagrammatic form,
using it to hide non-universal details
such as the complicated form of
the covariantization. However, before doing this,
we will provide an explicit example of
 a valid
covariantization, for completeness. Nevertheless, it
should be understood that this is just one of an 
infinite number of choices which we can make but
which, in practice, we never do: in actual calculations, 
we implicitly
work with an infinite number of flow equations.
The reason we can do this is that 
there exists a powerful diagrammatic calculus
which enables us to perform general computations,
almost entirely at the diagrammatic 
level~\cite{aprop,mgiuc,evalues}, in a way such that
nonuniversal details cancel out.
Indeed, this
calculus has been employed in pure $\SU(N)$ Yang-Mills
to give a diagrammatic expression for the $\beta$-function,
from which the universal answer (at least at one and
two loops) can be directly extracted. This formula
can be trivially adapted to QCD and we will use
this to perform a very compact computation of the one-loop
$\beta$-function.

The outline of this paper is as follows. In \sec{YM},
we review the regularization of $\SU(N)$ Yang-Mills
via $\SU(N|N)$ Yang-Mills. In \sec{Quarks} we add
the quarks, first seeing why we cannot embed them
in the fundamental of $\SU(N|N)$ and then describing
how we can instead embed them using a more elaborate
scheme. We conclude this section by showing how
to give the quarks independent masses. In \sec{flow}
we review the construction of a manifestly
gauge invariant flow equation for $\SU(N)$ Yang-Mills
and then adapt it for QCD.
In \sec{beta1} we give a diagrammatic expression
for the one-loop $\beta$ function which reproduces
the universal result in the case that the quarks
are massless. Finally, in \sec{conc}, we conclude.

\section{Regularizing $\SU(N)$ Yang-Mills}
\label{sec:YM}

\subsection{Embedding in $\SU(N|N)$ Yang-Mills}

Throughout this paper, we work in Euclidean dimension, D.
We regularize $\SU(N)$ Yang-Mills by embedding
it in spontaneously broken $\SU(N|N)$ Yang-Mills,
which is itself regularized by covariant higher derivatives~\cite{SU(N|N)}.
The supergauge field, $\A_\mu$, is valued in
the Lie superalgebra and, using the defining representation,
can be written as a Hermitian supertraceless supermatrix
(the supertrace of a supermatrix is defined as the trace
of the top block diagonal element minus the trace
of the bottom block diagonal element):
\be
\label{eq:defA}
	\A_\mu = 
	\left(
		\begin{array}{cc}
			A_\mu^1 	& B_\mu 
		\\
			\bar{B}_\mu & A_\mu^2
		\end{array} 
	\right) + \A_\mu^0 \one.
\ee
Here, $A^1_\mu(x)\equiv A^1_{a\mu}\tau^a_1$ is the
physical $\SU(N)$ gauge field, $\tau^a_1$ being the $\SU(N)$
generators orthonormalized to
$\tr(\tau^a_1\tau^b_1)=\delta^{ab}/2$, while $A^2_\mu(x)\equiv
A^2_{a\mu}\tau^a_2$ is a second, unphysical $\SU(N)$ gauge field.
The $B$ fields are fermionic gauge fields which will gain a mass
of order $\Lambda$ from the spontaneous symmetry breaking; they play the
role of gauge invariant PV fields, furnishing the
necessary extra regularization to supplement the covariant
higher derivatives. In order
to unambiguously define contributions which are finite
only by virtue of the PV regularization, a preregulator must
be used in $D=4$~\cite{SU(N|N)}. We will use
dimensional regularization, emphasising that
this makes sense
non-perturbatively, since 
it is not being used to renormalize the theory, but rather
as a prescription for discarding surface terms in loop
integrals~\cite{SU(N|N),ym}.

$\A^0$ is the gauge field for the centre of the
$\SU(N|N)$ Lie superalgebra. Equivalently, one can write
\be
\label{eq:expandA}
\A_\mu = {\A}^{0}_{\mu} \one + \A^A_\mu T_A,
\ee
where the $T_A$ are a complete set of traceless and supertraceless
generators normalised as in~\cite{SU(N|N)}.

The theory is
subject to the local invariance:
\be
\label{eq:Agauged}
\delta\A_\mu = [\nabla_\mu,\Omega(x)] +\lambda_\mu(x) \one.
\ee
The first term, in which $\nabla_\mu = \partial_\mu -i\A_\mu$, 
generates supergauge transformations. Note that the coupling, $g$,
has been scaled out of this definition. It is worth doing
this: since we do not gauge fix, the exact preservation of~\eq{Agauged}
means that none of the fields suffer wavefunction
renormalization, even in the broken phase~\cite{aprop}.
The second term in~\eq{Agauged} divides out the centre of the algebra.
The reason for doing this is as follows. The superfield
strength is $\F_{\mu\nu} = i[\nabla_\mu,\nabla_\nu]$, out
of which we construct the kinetic term  $\sim \str \F_{\mu\nu}^2$.
On account of $\str \one = 0$ and the fact that $\one$
commutes with everything, it is apparent that 
$\A^0$ has neither a kinetic term, nor any interactions.
Consequently, if $\A^0$
were to appear anywhere else in the action, it would act as a Lagrange
multiplier and so we forbid its presence. The resulting
`no-$\A^0$ shift symmetry' ensures that nothing depends on $\A^0$
and that $\A^0$ has no degrees of freedom.%
\footnote{
It is tempting to try and remove $\A^0$ from the algebra. However,
we cannot do this directly since although
$\SU(N|N)$ is reducible it is indecomposable: $\A^0$ is generated
by (fermionic) gauge transformations. It is possible to instead
modify the Lie bracket~\cite{Bars} but this appears only
to complicate matters.
}
This will prove important
when we come to add quarks.

The spontaneous breaking is carried by a superscalar
field
\[
\C =
	\left(
		\begin{array}{cc}
			C^1		& D
		\\
			\bar{D}	& C^2
		\end{array}
	\right).
\]
This field is Hermitian but, unlike
$\A_\mu$, is not supertraceless
and so it is valued in the $\U(N|N)$ Lie algebra.
Nevertheless, the whole of $\C$ transforms
homogeneously under local $\SU(N|N)$:
\be
\label{eq:Cgauged}
\delta\C = -i\,[\C,\Omega].
\ee

It can be shown that, at the classical level, the spontaneous
breaking scale (effectively the mass of $B$) tracks the covariant
higher derivative effective cutoff scale, $\Lambda$, if $\C$ is
made dimensionless (by using powers of $\Lambda$) and  $\hS$ has
the minimum of its effective potential at:
\be
\label{eq:sigma}
\expectation{\C} = \sigma \equiv \pmatrix{\one_N & 0\cr 0 & -\one_N},
\ee
where $\one_N$ is the $N\times N$ identity matrix.

In this case the classical action $S_0$ also has a minimum 
at~\eq{sigma}. At the quantum level this can be imposed as a
constraint on $S$ by taking $\expectation{\C} = \sigma $ as a renormalization
condition. This ensures that the Wilsonian effective action
does not possess any one-point vertices, which can be
translated into a constraint on
$\hS$~\cite{aprop,Thesis}. In the broken phase, $D$ is
a super-Goldstone mode (eaten by $B$ in the unitary gauge) whilst the
$C^i$ are Higgs bosons and can be given a running mass of order
$\Lambda$~\cite{ym,SU(N|N),aprop}. Working in our manifestly
gauge invariant formalism, $B$ and $D$ gauge transform into each
other, as we will see in \sec{YM-WIDs}.

In addition to the coupling, $g$, of the physical gauge
field, the field $A^2_\mu$ carries its own coupling, $g_2$,
(in the broken phase)
which renormalizes separately~\cite{aprop,Thesis,mgierg1,mgierg2}.
It is often useful not to work with $g_2$ directly but rather
with
\be
	\alpha \equiv g^2_2/g^2.
\label{eq:alpha-defn}
\ee
The couplings $g$ and $\alpha$ 
are defined through their
renormalization conditions:
\bea
\label{eq:defg}
	S[\A=A^1, \C=\sigma]	& =	& {1\over2g^2}\,\str\!\int\!\!d^D\!x\,
									\left(F^1_{\mu\nu}\right)^2+\cdots,
\\	
\label{eq:defg2}
	S[\A=A^2, \C=\sigma] 	& =	& {1\over2 \alpha g^2}\,\str\!\int\!\!d^D\!x\,
									\left(F^2_{\mu\nu}\right)^2+\cdots,
\eea
where the ellipses stand for higher dimension operators and the
ignored vacuum energy. The field strength tensors in the $A^1$ and
$A^2$ sectors, $F^1_{\mu\nu}$ and $F^2_{\mu\nu}$, should really
be embedded in the top left / bottom right entries of a supermatrix,
in order for the supertraces in~\eqs{defg}{defg2} 
to make sense. We will frequently employ
this minor abuse of notation, for convenience.

\subsection{Ward Identities}
\label{sec:YM-WIDs}

The supergauge invariant Wilsonian effective action
has an expansion in terms of supertraces and
products of supertraces~\cite{aprop}:
\begin{eqnarray}
\fl
	S 	& =	& \sum_{n=1}^\infty \frac{1}{s_n} \Int{x_1} \!\! \cdots \volume{x_n} 
				S^{X_1 \cdots X_n}_{\,a_1 \, \cdots \, a_n}(x_1, \cdots, x_n) \str X_1^{a_1}(x_1) \cdots X_n^{a_n} (x_n)
	\nonumber \\
\fl		& +	& \frac{1}{2!} \sum_{m,n=0}^{\infty} \frac{1}{s_n s_m} \Int{x_1} \!\!\cdots \volume{x_n}  \volume{y_1} \!\! \cdots \volume{y_m}
				S^{X_1 \cdots X_n, Y_1 \cdots Y_m}_{\, a_1 \, \cdots \, a_n \, , \, b_1\cdots \, b_m}(x_1, \cdots, x_n; y_1 \cdots y_m)
	\nonumber \\
\fl		&	&	
			\qquad \str X_1^{a_1}(x_1) \cdots X_n^{a_n} (x_n) \, \str Y_1^{b_1}(y_1) \cdots Y_m^{b_m} (y_m)
	\nonumber \\
\fl		& + & \ldots 
\label{eq:YMActionExpansion}
\end{eqnarray}
where the $X^{a_i}_i$ 
and $Y^{b_j}_j$ are any of the broken phase fields, with the $a_i$
and $b_j$ being Lorentz indices or null, as appropriate.
The vacuum energy is ignored. We take only one cyclic ordering for the 
lists $X_1 \cdots X_n$, $Y_1 \cdots Y_m$ in the sums over $n,m$.
If any term is invariant under some nontrivial cyclic
permutations of its arguments, then $s_n$ ($s_m$) is the order of the
cyclic subgroup, otherwise $s_n = 1$ ($s_m = 1$).

The momentum space vertices are written
\[
\fl
	S^{X_1 \cdots X_n}_{\,a_1 \, \cdots \, a_n}(p_1, \cdots, p_n) \left(2\pi \right)^D \delta \left(\sum_{i=1}^n p_i \right)
	=
	\Int{x_1} \!\! \cdots \volume{x_n} e^{-i \sum_i x_i \cdot p_i} S^{X_1 \cdots X_n}_{\,a_1 \, \cdots \, a_n}(x_1, \cdots, x_n),
\]
where all momenta are taken to point into the vertex. We employ the shorthand
\[
	S^{X_1 X_2}_{\, a_1 \; a_2}(p) \equiv S^{X_1 X_2}_{\, a_1 \; a_2}(p,-p).
\]

Since we will ultimately be giving the
flow equation for QCD via its diagrammatic representation,
it is useful at this stage to introduce diagrammatics
for the action~\eq{YMActionExpansion}.
The \emph{vertex coefficient functions} belonging to the
action~\eq{YMActionExpansion} have a simple diagrammatic
representation: 
\be
	\dec{
		\ensuremath{\begin{array}{c}\begin{picture}(0,0)%
\includegraphics{pstex/Vertex-S.pstex}%
\end{picture}%
\setlength{\unitlength}{3947sp}%
\begingroup\makeatletter\ifx\SetFigFont\undefined%
\gdef\SetFigFont#1#2#3#4#5{%
  \reset@font\fontsize{#1}{#2pt}%
  \fontfamily{#3}\fontseries{#4}\fontshape{#5}%
  \selectfont}%
\fi\endgroup%
\begin{picture}(341,318)(2180,-963)
\put(2291,-859){\makebox(0,0)[lb]{\smash{{\SetFigFont{11}{13.2}{\rmdefault}{\mddefault}{\updefault}{\color[rgb]{0,0,0}$S$}%
}}}}
\end{picture}%
 \end{array}}
	}{\{f\}}
	\equiv
	\ensuremath{\begin{array}{c}\begin{picture}(0,0)%
\includegraphics{pstex/Vertex-S-f.pstex}%
\end{picture}%
\setlength{\unitlength}{3947sp}%
\begingroup\makeatletter\ifx\SetFigFont\undefined%
\gdef\SetFigFont#1#2#3#4#5{%
  \reset@font\fontsize{#1}{#2pt}%
  \fontfamily{#3}\fontseries{#4}\fontshape{#5}%
  \selectfont}%
\fi\endgroup%
\begin{picture}(395,455)(2180,-963)
\put(2291,-859){\makebox(0,0)[lb]{\smash{{\SetFigFont{11}{13.2}{\rmdefault}{\mddefault}{\updefault}{\color[rgb]{0,0,0}$S$}%
}}}}
\end{picture}%
 \end{array}}
\label{eq:WEA-VCF}
\ee
represents all vertex coefficient functions corresponding
to all cyclically independent orderings of the
set of broken phase fields, $\{f\}$, distributed
over all possible supertrace structures. For example,
\be
\label{eq:Vertex-C1C1}
	\dec{
		\ensuremath{\begin{array}{c}\begin{picture}(0,0)%
\includegraphics{pstex/Vertex-S.pstex}%
\end{picture}%
\setlength{\unitlength}{3947sp}%
\begingroup\makeatletter\ifx\SetFigFont\undefined%
\gdef\SetFigFont#1#2#3#4#5{%
  \reset@font\fontsize{#1}{#2pt}%
  \fontfamily{#3}\fontseries{#4}\fontshape{#5}%
  \selectfont}%
\fi\endgroup%
\begin{picture}(341,318)(2180,-963)
\put(2291,-859){\makebox(0,0)[lb]{\smash{{\SetFigFont{11}{13.2}{\rmdefault}{\mddefault}{\updefault}{\color[rgb]{0,0,0}$S$}%
}}}}
\end{picture}%
 \end{array}}
	}{C^1C^1}
\ee
represents both the coefficient functions $S^{C^1 C^1}$ and
$S^{C^1,C^1}$ which, from~\eq{YMActionExpansion}, 
are associated with the
supertrace structures $\str C^1 C^1$ 
and $\str C^1 \str C^1$, respectively.
(We have suppressed the momentum arguments.)
Similarly,
\[
	\left[
		\ensuremath{\begin{array}{c}\begin{picture}(0,0)%
\includegraphics{pstex/Vertex-S.pstex}%
\end{picture}%
\setlength{\unitlength}{3947sp}%
\begingroup\makeatletter\ifx\SetFigFont\undefined%
\gdef\SetFigFont#1#2#3#4#5{%
  \reset@font\fontsize{#1}{#2pt}%
  \fontfamily{#3}\fontseries{#4}\fontshape{#5}%
  \selectfont}%
\fi\endgroup%
\begin{picture}(341,318)(2180,-963)
\put(2291,-859){\makebox(0,0)[lb]{\smash{{\SetFigFont{11}{13.2}{\rmdefault}{\mddefault}{\updefault}{\color[rgb]{0,0,0}$S$}%
}}}}
\end{picture}%
 \end{array}}
	\right]^{A^1A^1C^1}_{\mu \ \; \nu}
\]
represents $S^{A^1 A^1 C^1}_{\mu \ \; \nu}$, 
$S^{A^1 A^1 C^1}_{\nu \ \; \mu}$
and $S^{A^1 A^1, C^1}_{\mu \ \; \nu}$. (There are
no vertices which correspond to a trace of
a single $A^1$, since $\str A^1 = 0$.)

The (un)broken gauge transformations
follow from splitting $\Omega$ into
its block components
\[
	\Omega = 
		\left(
		\begin{array}{cc}
			\omega^1	& \tau
		\\
			\bar{\tau}	& \omega^2
		\end{array}
	\right) + \Omega^0 \one
\]
and expanding out~\eqs{Agauged}{Cgauged} 
(we are not interested in the no-$\A^0$ symmetry, here). For this purpose,
it is useful to combine the fields $A^1$ and $A^2$
with the block diagonal components of $\A_0 \one$.
We denote the resultant fields by $\tilde{A}^1$
and $\tilde{A}^2$, though note that sometimes
$\A^0$ contributions can cancel out between terms.
This gives the
unbroken $\SU(N)\times\SU(N)\times U(1)$ transformations~\cite{aprop}
\be
\label{eq:YM-unbroken}
\fl
	\begin{array}{lcl}
		\delta \tilde{A}^1_\mu 	& =	& D^1_\mu \cdot \omega^1 + \partial_\mu \Omega^0 \one_N
	\\
		\delta B_\mu			& =	& -i(B_\mu \omega^2 - \omega^1 B_\mu)
	\\	
		\delta C^1				& =	& -i C^1 \cdot \omega^1
	\\	
		\delta D				& =	& -i (D \omega^2 - \omega^1 D)
	\end{array}
	\qquad
	\begin{array}{lcl}
		\delta \tilde{A}^2_\mu 	& =	& D^2_\mu \cdot \omega^2 + \partial_\mu \Omega^0 \one_N
	\\
		\delta \bar{B}_\mu		& =	& -i(\bar{B}_\mu \omega^1 - \omega^2 \bar{B}_\mu)
	\\	
		\delta C^2				& =	& -i C^2 \cdot \omega^2
	\\
		\delta \bar{D}			& =	& -i (\bar{D} \omega^1 - \omega^2 \bar{D})
	\end{array}
\ee
and the broken fermionic gauge transformations
\be
\label{eq:YM-broken}
\fl
	\begin{array}{lcl}
		\delta \tilde{A}^1_\mu 	& =	& -i(B_\mu \bar{\tau} - \tau \bar{B}_\mu)
	\\
		\delta B_\mu			& =	& \partial_\mu \tau - i(A^1_\mu \tau - \tau A^2_\mu)
	\\	
		\delta C^1				& =	& -i (D \bar{\tau} - \tau \bar{D})
	\\	
		\delta D				& =	& -i (C^1 \tau - \tau C^2) - 2i \tau
	\end{array}
	\qquad
	\begin{array}{lcl}
		\delta \tilde{A}^2_\mu 	& =	& -i(\bar{B}_\mu \tau - \bar{\tau} B_\mu)
	\\
		\delta \bar{B}_\mu		& =	& \partial_\mu \bar{\tau} - i(A^2_\mu \bar{\tau} - \bar{\tau} A^1_\mu)
	\\	
		\delta C^2				& =	& -i(\bar{D} \tau - \bar{\tau} D)
	\\
		\delta \bar{D}			& =	& - i(C^2 \bar{\tau} - \bar{\tau} C^1) +2i\bar{\tau},
	\end{array}
\ee
where $D^{(1,2)} = \partial_\mu -i A^{1,2}_\mu$
are the covariant derivatives appropriate to 
the physical gauge field and the unphysical copy
and the dot again means action by commutation.

As noted in~\cite{aprop}, the manifest
preservation of the transformations
for $\tilde{A}^1_\mu$ and $\tilde{A}^2_\mu$
in~\eq{YM-unbroken} protects these fields
from field strength renormalization.
The remaining fields are similarly protected,
as follows from~\eq{YM-broken}.

The transformations~\eq{YM-broken} for $B_\mu$ and $D$,
$\bar{B}_\mu$ and $\bar{D}$ leads us to define~\cite{aprop}%
\footnote{Actually, these definitions differ from
those of~\cite{aprop} by a sign in the fifth component.
They are, however, consistent with~\cite{Thesis,mgierg1,mgierg2,RG2005,Primer,mgiuc,evalues}.
}
\numparts
\bea
\label{eq:F}
	F_M & = & (B_\mu, D),
\\
\label{eq:Fbar}
	\bar{F}_N & = & (\bar{B}_\nu, -\bar{D}),
\eea
\endnumparts
where $M$, $N$ are
five-indices~\cite{Thesis,mgierg1}. The summation
convention for these indices is that we take each product of
components to contribute with unit weight.

Two Ward identities now follow from applying~\eqs{YM-unbroken}{YM-broken}
to the action~\eq{YMActionExpansion}.  The transformations~\eq{YM-unbroken}
yield:
\be
\label{eq:YM-WID-unbroken}
\fl
	q_\nu S^{\cdots X \! A^{1,2} Y \cdots}_{\cdots \, a \; \nu \hspace{0.9em} b \hspace{-0.1em} \ \cdots} (\ldots, p,q,r, \ldots) 
	=
	S^{\cdots XY \cdots}_{\cdots \, a \; b \ \cdots}( \ldots, p, q+r, \ldots) - 	
	S^{\cdots XY \cdots}_{\cdots \, a \; b \ \cdots}( \ldots, p+q, r, \ldots).
\ee
The effect of the transformations~\eq{YM-broken}
are most efficiently written in the five component
language of~\eqs{F}{Fbar}.
Introducing a five momentum
\be
	q_M = (q_\mu, 2),
\label{eq:q}
\ee
allows us to write
\be
\label{eq:YM-WID-broken}
\fl
	q_N S^{\cdots X \! F Y \cdots}_{\cdots \, a  N  b \ \cdots} (\ldots, p,q,r, \ldots) 
	=
	S^{\cdots X\PushF{Y} \cdots}_{\cdots \, a \; b \ \cdots}( \ldots, p, q+r, \ldots) - 	
	S^{\cdots \PullB{X}Y \cdots}_{\cdots \, a \; b \ \cdots}( \ldots, p+q, r, \ldots),
\ee
where $\PushF{Y}$ and $\PullB{X}$ are the opposite statistics
partners of the fields $Y$ and $X$. (For explicit expressions
see~\cite{Thesis,mgierg1}.) An identical expression 
to~\eq{YM-WID-broken} exists for when the field $F_N$ is replaced
by $\bar{F}_N$.

The Ward identities~\eqs{YM-WID-unbroken}{YM-WID-broken}
can be beautifully combined using the diagrammatics:
\be
\label{eq:WID-A}
\fl
	\ensuremath{\begin{array}{c}\input{pstex/WID-contract.pstex_t} \end{array}} = \ensuremath{\begin{array}{c}\input{pstex/WID-PF.pstex_t} \end{array}} + \ensuremath{\begin{array}{c}\input{pstex/WID-PFb.pstex_t} \end{array}} - \ensuremath{\begin{array}{c}\input{pstex/WID-PB.pstex_t} \end{array}} - \ensuremath{\begin{array}{c}\input{pstex/WID-PBb.pstex_t} \end{array}} + \cdots
\ee

On the \lhs, we contract a vertex with the momentum of
the field which carries $p$. This field---which we will
call the active field---can be either
$A^1_\rho$, $A^2_\rho$, $F_R$ or $\bar{F}_R$.
In the first two cases, the triangle $\GRk$ represents
$p_\rho$ whereas, in the latter two cases, it represents
$p_R = (p_\rho,2)$. (Given that we often sum over
all possible fields, we can take the Feynman rule for
$\GRk$ in the $C$-sector to be null.)
On the \rhs\ of~\eq{WID-A}, we push the contracted momentum forward onto 
the field which directly follows the active field, in the counterclockwise
sense, and pull back (with a minus sign) onto
the field which directly precedes the active field. 
Since our diagrammatics is permutation symmetric, the struck field---which
we will call the target field---can
be either $X$, $Y$ or any of the un-drawn fields, 
as represented by the ellipsis.

Allowing the active
field to strike another field necessarily involves a partial
specification of the supertrace structure: it must be the case that
the struck field either directly followed or preceded the active
field. In turn, this means that the Feynman rule for particular
choices of the active and target fields can be zero. For example,
as trivially follows by multiplying together supermatrices,
an $F$ can follow, but never precede an $A^1_\mu$, and so the 
pull back of an $A^1_\mu$ onto an $F$ should be assigned a value
of zero. 
The momentum routing follows in an obvious manner: for example,
in the first diagram on the \rhs, momenta $q+p$ and $r$ now flow into
the vertex. In the case that the active field is fermionic,
the field pushed forward / pulled back onto is transformed
into its opposite statistic partner, as above. 

The half arrow which terminates the pushed forward / pulled back
active field is of no significance and can go on either side
of the active field line. It is necessary to
keep the active field line---even though the active field
is no longer part of the vertex---in order that
we can unambiguously deduce flavour changes 
and momentum routing, without reference to the parent diagram.

We illustrate~\eq{WID-A} by considering contracting
$\GRk$ into the Wilsonian the effective action
two-point vertex:
\be
\label{eq:GR-TP}
	\ensuremath{\begin{array}{c}\begin{picture}(0,0)%
\includegraphics{pstex/GR-TP.pstex}%
\end{picture}%
\setlength{\unitlength}{3947sp}%
\begingroup\makeatletter\ifx\SetFigFont\undefined%
\gdef\SetFigFont#1#2#3#4#5{%
  \reset@font\fontsize{#1}{#2pt}%
  \fontfamily{#3}\fontseries{#4}\fontshape{#5}%
  \selectfont}%
\fi\endgroup%
\begin{picture}(757,318)(1880,-963)
\put(2278,-861){\makebox(0,0)[lb]{\smash{{\SetFigFont{11}{13.2}{\rmdefault}{\mddefault}{\updefault}{\color[rgb]{0,0,0}$S$}%
}}}}
\end{picture}%
 \end{array}} =  \ensuremath{\begin{array}{c}\begin{picture}(0,0)%
\includegraphics{pstex/GR-TP-PF.pstex}%
\end{picture}%
\setlength{\unitlength}{3947sp}%
\begingroup\makeatletter\ifx\SetFigFont\undefined%
\gdef\SetFigFont#1#2#3#4#5{%
  \reset@font\fontsize{#1}{#2pt}%
  \fontfamily{#3}\fontseries{#4}\fontshape{#5}%
  \selectfont}%
\fi\endgroup%
\begin{picture}(457,414)(2180,-1059)
\put(2278,-861){\makebox(0,0)[lb]{\smash{{\SetFigFont{11}{13.2}{\rmdefault}{\mddefault}{\updefault}{\color[rgb]{0,0,0}$S$}%
}}}}
\end{picture}%
 \end{array}} - \ensuremath{\begin{array}{c}\begin{picture}(0,0)%
\includegraphics{pstex/GR-TP-PB.pstex}%
\end{picture}%
\setlength{\unitlength}{3947sp}%
\begingroup\makeatletter\ifx\SetFigFont\undefined%
\gdef\SetFigFont#1#2#3#4#5{%
  \reset@font\fontsize{#1}{#2pt}%
  \fontfamily{#3}\fontseries{#4}\fontshape{#5}%
  \selectfont}%
\fi\endgroup%
\begin{picture}(457,414)(2180,-963)
\put(2278,-861){\makebox(0,0)[lb]{\smash{{\SetFigFont{11}{13.2}{\rmdefault}{\mddefault}{\updefault}{\color[rgb]{0,0,0}$S$}%
}}}}
\end{picture}%
 \end{array}}.
\ee
Given that $\GRk$ is null in the $C^i$ sector,
the fields decorating the two-point
vertex on the \rhs\ can be either both $A^i$s
or both fermionic. In the former case, 
\eq{GR-TP} reads:
\[
	p_\mu S^{A^i A^i}_{\mu \ \, \nu}(p) = S^{A^i}_{\nu}(0) - S^{A^i}_{\nu}(0) = 0
\]
where we note that $S^{A^i}_{\nu}$ is in fact zero by itself,
as follows by both Lorentz invariance and gauge invariance.
In the latter case, \eq{GR-TP} reads:
\[
	p_M S^{\bar{F} \; F}_{M N}(p) = \left[S^{C^2}(0) - S^{C^1}(0)\right] \delta_{N5},
\]
where we have used~\eq{F} and have discarded
contributions which go like $S^{A^i}_{\nu}(0)$.
However, the $S^{C^i}(0)$ must vanish. This
follows from demanding
that the minimum of the superhiggs potential is
not shifted by quantum corrections~\cite{aprop}.
Therefore,
\be
\label{eq:D-ID-GR-TP-A}
	\ensuremath{\begin{array}{c}\begin{picture}(0,0)%
\includegraphics{pstex/GR-TP.pstex}%
\end{picture}%
\setlength{\unitlength}{3947sp}%
\begingroup\makeatletter\ifx\SetFigFont\undefined%
\gdef\SetFigFont#1#2#3#4#5{%
  \reset@font\fontsize{#1}{#2pt}%
  \fontfamily{#3}\fontseries{#4}\fontshape{#5}%
  \selectfont}%
\fi\endgroup%
\begin{picture}(757,318)(1880,-963)
\put(2278,-861){\makebox(0,0)[lb]{\smash{{\SetFigFont{11}{13.2}{\rmdefault}{\mddefault}{\updefault}{\color[rgb]{0,0,0}$S$}%
}}}}
\end{picture}%
 \end{array}} = 0.
\ee

\subsection{Taylor Expansion of Vertices}
\label{sec:Taylor}

For the formalism to be properly defined,
it must be the case that all vertices
are Taylor expandable to all orders
in momenta~\cite{ym,ymi,ymii}.
Consider a vertex which is part
of a complete diagram, decorated by some set of internal
fields and by a single external $A^1$ (or $A^2$), which we
denote by a wiggly line.
The diagrammatic representation for the zeroth order expansion
in the momentum of the external field is all that is required
for this paper~\cite{Thesis,mgierg1}:
\be
\label{eq:Taylor-A}
\fl
	\ensuremath{\begin{array}{c}\input{pstex/Taylor-Parent.pstex_t} \end{array}} = \cdeps{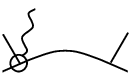} + \cdeps{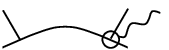} - \cdeps{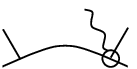} - \cdeps{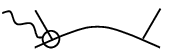} +\cdots;
\ee
note the similarity to~\eq{WID-A}.

The interpretation of the diagrammatics is as follows. In the first diagram
on the \rhs, the vertex is differentiated \wrt\ the momentum carried
by the field $X$, whilst holding the momentum of the preceding field fixed
(we assume for the time being that both $X$ and the
preceding field carry non-zero momentum).
Of course, using our current diagrammatic notation,
this latter field can be any of those
which decorate the vertex, and so we sum over all possibilities. Thus,
each cyclically ordered push forward like term has a partner,
cyclically ordered pull back like term, such that
the pair can be interpreted as
\be
	\left( \left. \partial^r_\mu \right|_s - \left. \partial^s_\mu \right|_r \right) \mathrm{Vertex},
\label{eq:Momderivs}
\ee
where $r$ and $s$ are momenta entering the vertex. 
In the case that $r=-s$, we can and will
drop either the push forward like term or pull back like term, since
the combination can be expressed as $\partial^r_\mu$; we
interpret the diagrammatic notation appropriately.
If any of the fields decorating the vertex carry
zero momentum (besides the explicitly drawn $A^i$),
then they are transparent to this entire procedure.
Thus, they are never differentiated and, if they precede
a field which is, we must look to the first field carrying
non-zero momentum to figure
out which of the vertex's momenta is held constant.

\section{Adding Quarks}
\label{sec:Quarks}

\subsection{Massless Quarks}

The simplest way to try to incorporate quarks into the setup
is to embed them into the fundamental representation of
$\SU(N|N)$:
\[
	\Psi = 
	\left(
	\begin{array}{c}
		\psi
	\\
		\varphi
	\end{array}
	\right),
\]
where $\Psi$ transforms under $\SU(N|N)$, $\psi$ is a physical
quark field and $\varphi$ is an unphysical, bosonic 
spinor (here and henceforth, we suppress spinor indices).%
\footnote{
If this scheme were to work, we would also have to introduce further
unphysical fields to provide sufficient PV regularization.
}
Immediately, we can see that this embedding is inconsistent with~\eq{Agauged}:
the supergauge invariant quark term,
\[
	\frac{1}{g^2} i \bar{\Psi} \Delslash \Psi,
\]
does not satisfy no-$\A^0$ symmetry.

If, however, the fields $\Psi$ transform as $R \otimes \bar{R}$, for 
some representation, $R$, then we can construct a no-$\A^0$
invariant
representation simply because $\Psi$ has 
zero `charge' under no-$\A^0$.
Thus, the strategy we pursue is to embed the quarks
into fields which transform as a bifundamental 
of $\SU(N|N)$. To achieve
this, we first embed the up-like quarks [up, charm, top (suitably generalized
for $N\neq3$)]
and down-like quarks (down, strange, bottom) 
into two tensor fields $\left(\up\right)^i_j$ and $\left(\down\right)^i_j$,
where the superscript indices carry an $\SU(N)$ colour
symmetry and the subscript indices carry an (unphysical, gauged)
$\SU(N)$ flavour symmetry. In turn, $\up$ and $\down$ 
are now embedded into fields $\Up$ and $\Down$
which are valued in complexified $\U(N|N)$:
\be
\label{eq:Embed}
	\Up =
	\left(
		\begin{array}{cc}
			\varphi^1 	& \up
		\\
			\varrho		& \varphi^2
		\end{array}
	\right),
	\qquad
	\Down =
	\left(
		\begin{array}{cc}
			\phi^1 	& \down
		\\
			\rho	& \phi^2
		\end{array}
	\right).
\ee
Notice that $\Up$ and $\Down$ are not Hermitian. Consequently,
$\varrho$ ($\rho$) are not related to the physical
fields $\up$ ($\down$) and, since they will be seen to 
come with wrong
sign action, should be interpreted as
unphysical degrees of freedom. These fields,
together with $\phi^1$, $\phi^2$,
$\varphi^1$ and $\varphi^2$ (the components
of which are bosonic spinors), will be given a
mass of order the cutoff.
Under gauge transformations,
$\Up$ and $\Down$ transform homogeneously:
\be
\label{eq:Qgauged}
	\delta \Up = -i[\Up,\Omega], \qquad \delta \Down = -i[\Down,\Omega].
\ee

The $\SU(N|N)$ invariant quark kinetic term that 
we include in the Lagrangian is just
\be
\label{eq:QuarksLagranian}
	-\frac{i}{g^2} 
	\left( 
		\str \antiUp \Delslash \cdot \Up + \str \antiDown \Delslash \cdot \Down
	\right),
\ee
where the minus sign compensates the sign
buried in the supertrace, ensuring 
that the physical quark
terms come with the correct overall sign.
Notice that we have not included
any covariant higher derivatives; it is
straightforward to demonstrate that the supergroup
structure, alone, is sufficient to provide
the necessary regularization in the quark sector
by repeating the analysis of~\cite{SU(N|N)},
but this time including the fields $\Quarks_i$.

To show the types of terms that we must include in
the action to give the unphysical fields a mass
of order the cutoff but leave the quarks massless (for
the time being), it is useful to construct
the following projectors:
\be
\label{eq:sigma_pm}
	\sigma_{+} \equiv \frac{1}{2}(\one + \sigma) = 
	\left(
		\begin{array}{cc}
			\one_N	&	0
		\\
				0	&	0
		\end{array}
	\right), \qquad
		\sigma_{-} \equiv \frac{1}{2}(\one - \sigma) = 
	\left(
		\begin{array}{cc}
			0	&	0
		\\
			0	&	\one_N
		\end{array}
	\right).
\ee

With a slight abuse of notation, we can write
$\varphi^1 = \sigma_+ \Up \sigma_+$, $\varphi^2 = \sigma_- \Up \sigma_-$,
$\varrho = \sigma_- \Up \sigma_+$ and $\up = \sigma_+ \Up \sigma_-$.
We can lift these projectors to the symmetric phase by
defining
\be
\label{eq:SymPhasPro}
	\varsigma_\pm \equiv \frac{1}{2} (\one \pm \C).
\ee
Thus, to give a mass to \eg\ $\varrho$ in the broken phase, 
all we need to do is add to the Lagrangian the term:
\[
	-\frac{1}{g^2} \Lambda \str \left(\antiUp \varsigma_- \Up \varsigma_+ \right).
\]
Upon spontaneous symmetry breaking, this reduces to
$-\Lambda/ g^2 \tr \bar{\varrho} \varrho$ (plus interaction terms).

Thus, in the broken phase, the only massless fields
we are left with are $\quarks_i$, $A^1$ and $A^2$.
In the pure gauge case, we know from~\cite{SU(N|N)}
that $A^2$ decouples from $A^1$: integrating out
the heavy fields, the lowest dimension
gauge invariant effective interaction
left between $A^1$ and $A^2$ is the square of the two field
strengths (according to standard perturbative power
counting with $g$ in the usual place)
\[
	\Lambda^{-D} \tr (F^1)^2 \tr (F^2)^2,
\]
which is clearly irrelevant.
Adding the quarks, however, 
we see immediately from~\eq{QuarksLagranian}
that these fields can combine with the unphysical
gauge field to form a term of mass dimension $D$,
\[
\alpha \tr (\antiup \up + \antidown \down) \! \not{\! \!A^2}.
\]
To remove this term, and thus ensure that the unphysical
fields decouple as the regularization scale is sent
to infinity, we must tune $\alpha$ (see~\eq{alpha-defn}) 
to zero at the end
of a generic calculation. In fact, this
tuning is perfectly natural from a non-perturbative
perspective: the theory carried by $A^2$
is, as a consequence of its wrong sign action, not
asymptotically free but instead trivial. Moreover,
we can remove the need to perform any such tuning
by modifying the Higgs sector to completely
break the unphysical $\SU(N)$; indeed, we 
will do precisely this when we adapt the formalism
to give the quarks
independent masses.

We conclude this section by discussing 
the additional (un)broken invariances
which arise from the inclusion of the
quarks. The up-like quarks
supplement~\eq{YM-unbroken} with 
\be
\label{eq:Quarks-unbroken}
	\begin{array}{lcl}
		\delta \up			& =	& i (\omega^1 \up - \up \omega^2 )
	\\
		\delta \varrho		& =	& i(\omega^2 \varrho - \varrho \omega^1 )
	\\
		\delta \varphi^1	& =	& i \omega^1 \cdot \varphi^1	
	\\
		\delta \varphi^2	& =	& i \omega^2 \cdot \varphi^2
	\end{array}
\ee
and~\eq{YM-broken} with
\be
\label{eq:Quarks-broken}
	\begin{array}{lcl}
		\delta \up			& =	& i (\tau \varphi^2 - \varphi^1 \tau)
	\\	
		\delta \varrho		& =	& i(\bar{\tau} \varphi^1 - \varphi^2 \bar{\tau})
	\\
		\delta \varphi^1	& =	& i (\tau \varrho- \up \bar{\tau})
	\\	
		\delta \varphi^2	& =	& i(\bar{\tau} \up - \varrho \tau)
	\end{array}
\ee
(similarly for the down-like quarks). Notice
that the quark field $\up$ is not protected
from field strength renormalization. However,
the transformations~\eq{Quarks-broken} do 
enforce that all components of $\Up$ have
the same field strength renormalization
(likewise $\Down$).
The unbroken transformation for $\up$
given by~\eq{Quarks-unbroken} confirms
our interpretation 
that the physical colour
symmetry is carried by $A^1$, whereas the
unphysical flavour symmetry is carried by $A^2$.

\subsection{Massive Quarks}

If we only needed to give all the up-like  quarks 
one mass and all the down-like quarks one mass, 
then we could simply add a mass term
to the Lagrangian, using the fields that we have already:
\[
	-\frac{1}{g^2} 
	\left[
		m_u \, \str \antiUp \varsigma_+ \Up \varsigma_-
	+ 	m_d \, \str \antiDown \varsigma_+ \Down \varsigma_-
	\right],
\]
where we have used~\eq{SymPhasPro}.
Of course, to give all the quarks different
masses, we will have to break the unphysical, gauged
flavour symmetry.
To do this, we introduce two new
dimensionless
superscalars, $\C_u$ and $\C_d$
which, like $\C$, lie in the adjoint of
$\U(N|N)$ and  transform homogeneously:
\[
\delta\C_{u} = -i\,[\C_{u},\Omega], \qquad \delta\C_{d} = -i\,[\C_{d},\Omega].
\]
We choose the \vev\@s of $\C_u$ and $\C_d$
to be
\be
\label{eq:udvevs}
	\expectation{\C_u} =
	\left(
		\begin{array}{cc}
			\one_N	& 0
		\\
			0	& -\sigma_u
		\end{array}
	\right), \qquad
	\expectation{\C_d} =
	\left(
		\begin{array}{cc}
			\one_N	& 0
		\\
			0	& -\tilde{\sigma}_d
		\end{array}
	\right).
\ee
where $\sigma_u = \mathrm{diag} (1, m_c/m_u, m_t/m_u)$
and, given the unitary matrix, $U$,
$U^\dagger \tilde{\sigma}_d U =  \sigma_d = \mathrm{diag} (1, m_s/m_d, m_b/m_d)$
(with an obvious generalization for arbitrary values
of $N$).
The \vev\ of $\C_u$
breaks the unphysical $\SU(N)$ down to
$\U(1)^{N-1}$. We could choose the \vev\ of
$\C_d$ to be diagonal. However, in this
case we would be forced to tune the couplings
of the residual $\U(1)$s (which we note are, 
on account of their wrong sign action, asymptotically free
and not trivial) to zero. Consequently,
we  might as well choose
the \vev\ of $\C_d$ such that in
combination with the \vev\ of $\C_u$
the unphysical $\SU(N)$ is completely broken.
Thus, we 
implicitly assume $U$ to be such that 
amongst the generators broken
by $\expectation{\C_d}$ are those which are not broken
by $\expectation{\C_u}$. 
The \vev\@s of $\C_u$ and $\C_d$ (unlike that of $\C$)
are not protected from quantum corrections, which of
course corresponds to renormalization of the quark
masses. The broken Ward identities (whose modification
due to the breaking of the unphysical $\SU(N)$ we will
discuss shortly) protect the components of $\C_{u,d}$
from field strength renormalization, in the broken phase.

With the introduction of $\C_{u}$ and $\C_d$, there
is no requirement to retain $\C$. However, the following
exposition is made simpler if we keep $\C$ and so we
do so, noting that such considerations are
anyway irrelevant from the point of view of the diagrammatic
form of the flow equation.
We will not give an explicit realization
of the symmetry breaking
potential $V(\C,\C_u,\C_d)$ which yields~\eqs{sigma}{udvevs},
since we are free to work with any potential which 
satisfies the
following requirements. First,
(in unitarity gauge) all Goldstone bosons
are eaten by the various components
of $\A_\mu$ which acquire mass (\ie,
the potential must not possess any accidental
symmetries: the largest continuous 
symmetry group is just $\SU(N|N)$).
Secondly, the
remaining (Higgs) components of $\C$, $\C_u$ and $\C_d$
are given a mass of order the cutoff.

For non-degenerate masses, it is useful (and always possible) to  construct
the set of $N$ projectors which live in
the bottom right block
of a supermatrix:
\bea
	P_1 & = & \diag(0_N,1,0,0,\ldots)
\nonumber
\\
	P_2 & = & \diag(0_N,0,1,0,\ldots)
\nonumber
\\
	\vdots &&
\label{eq:BrokenPhaseProjectors}
\eea
To see this, simply note that \eg\
\[
	\frac{(\expectation{\C_u} - \expectation{\C})(\expectation{\C_u} - m_c/m_u\expectation{\C})}%
 		{(1-m_t/m_u)(m_c/m_u - m_t/m_u)}
	= \diag(0, 0, 0 ,0, 0, 1).  
\]
We can lift the $P_i$ to the symmetric phase
by introducing the non-degenerate (running)
parameters $a_i$ and defining:
\be
	\P_j = \prod_{i\neq j} \frac{\C_u - a_i \C}{a_i - a_j}.
\label{eq:Projectors}
\ee
In the broken phase (which recall that, by construction,
we are actually always in) we identify the $a_i$ with
the elements of $\sigma_u$. Note that the $\P_j$ 
gauge transform homogeneously.

The quarks' mass term can be taken to be
\be
\label{eq:QuarkMassTerm}
	\frac{1}{g^2}
	\left[
		m_u \str (\C_u \varsigma_- \antiUp \varsigma_+ \Up \varsigma_-)
		+m_d\str (\C_d \varsigma_- \antiDown \varsigma_+ \Down \varsigma_-)
	\right],
\ee
where it is understood from now on that we have rotated
the down-like quark fields to the mass basis
(this is exactly analogous to the introduction
of the CKM matrix in the standard model).
The remaining components of $\Psi_{u,d}$
are given masses of order the cutoff. Note that
we have included $\varsigma_\pm$ in~\eq{QuarkMassTerm}
purely for convenience, to ensure that the masses
of the components of the fields $\phi^2$
and $\varphi^2$ (see~\eq{Embed}) 
do not pick up contributions
from the quarks' mass matrices.

Neglecting the covariant higher derivative regularization,
the kinetic terms for $\C_u$ and $\C_d$
take the form
\[
	\frac{1}{2g^2} \str \!
	\left[
		(\nabla_\mu \cdot \C_u )^2 
		+ (\nabla_\mu \cdot \C_d)^2
	\right].
\]
Notice that this term provides 
differing contributions to the
masses 
of the various components of $B$.
Specifically, $B$ decomposes into
columns, with each column receiving
a different mass. This is precisely
what we would expect from the unbroken
gauge transformations, as we now
discuss.

An immediate effect of breaking the unphysical
$\SU(N)$ is that the relationships~\eqs{YM-unbroken}{YM-broken},
\eqs{Quarks-unbroken}{Quarks-broken}
(which we supplement by
those appropriate for $\C_{u,d}$)
decompose. 
The only relationships
which are completely unaffected are those
involving just $\omega_1$ \ie\
the unbroken relationships for $A^1_\mu$, $C^1$, $C_{u,d}^1$,
$\varphi^1$ and $\phi^1$.

The relationships involving just $\omega_2$
are completely broken. This means that the 
independent components of each of
the bottom right block fields are no longer
related by unbroken gauge transformations
and so can be expected to propagate separately.

In the fermionic sectors, the previously
unbroken gauge transformations involve
both $\omega_1$ and $\omega_2$ \eg\
$\delta B_\mu = -i(B_\mu \omega^2 - \omega^1 B_\mu)$.
Now, however, the $\omega_2$ part is completely broken.
In matrix language, the surviving unbroken 
transformation involving $\omega_1$
mixes up elements of each column with elements of
the \emph{same} column. Consequently, upon the
breaking of the unphysical $\SU(N)$, $B_\mu$
decomposes into $N$ `flavours', $B_{\mu a}$,
corresponding to the $N$ columns, with unbroken
transformation law 
\[
\bindexed{\delta B_{\mu a}}{i}{} = 
	i \bindexed{\omega^1}{i}{\;j} \bindexed{B_{\mu a}}{j}{}.
\]
This is precisely
what we want: in the quark sector, colour
remains a good symmetry and the unphysical,
gauged flavour symmetry
is completely broken.

With the above decomposition of many of our
fields, we must adapt our 
expansion of the action
in terms of fields~\eq{YMActionExpansion}
by appropriately expanding 
the set of fields represented by $X$ and $Y$.
To maintain the
supermatrix structure, we should ensure
that the new fields are still embedded
in supermatrices. For example, 
the field $B_{\mu a}$ should be in the appropriate
column of the top-right block of a supermatrix,
with all other elements set to zero. Equivalently,
we can project the field $B_{\mu a}$ out by
using the $P_i$ of~\eq{BrokenPhaseProjectors}.

At first sight, the breaking of the 
unphysical $\SU(N)$ considerably
complicates the Ward identities.
However, we can anticipate 
from~\cite{Primer,RG2005,mgiuc}
that we can and should hide these
complications in the diagrammatics.
This will be made particularly
straightforward if
we now sum over the flavours of the
target fields in~\eq{WID-A}. This helps
for the following reason. 
Consider~\eq{YM-WID-unbroken}
where  the
target fields are fermionic. We know that
the fermionic fields decompose by column
and so the \rhs\ of~\eq{YM-WID-unbroken}
will contain a sum of terms such that,
if the unphysical $\SU(N)$ were restored,
these terms could be combined back into
block supermatrix components.

We conclude this section by giving the renormalization
conditions for the quarks:
\[
\fl
	S=
		\frac{1}{g^2} \Int{x} 
		\tr 
		\left[ 
			\up \left(i\Delslash^1 +\sigma_u\right) \up + \down \left(i\Delslash^1 +\sigma_d\right)\down 
		\right] + \cdots,
\]
where the ellipsis represents all other operators
contributing to the effective action.

\section{A flow Equation for QCD}
\label{sec:flow}

\subsection{Review of $\SU(N)$ Yang-Mills}
\label{sec:flowreview}

\subsubsection{Setup}
\label{sec:YM-setup}

We begin by describing the flow equation used for
pure Yang-Mills~\cite{mgierg1}, the basic form of which is:
\be
\label{eq:flow}
	-\flow S = a_0[S,\Sigma_g] - a_1[\Sigma_g],
\ee
where $\Sigma_g \equiv g^2 S - 2\hat{S}$
($\hS$, we recall, being the seed action).%
\footnote{
Notice that we use $\Sigma_g$ instead of the $\Sigma_1$
of~\eq{Psi_x}, on account of $g$ being scaled out of
the covariant derivative.
}
On the right hand side
of the flow equation is the bilinear functional, $a_0[S,\Sigma_g]$,
which generates classical corrections and the functional
$a_1[\Sigma_g]$ which generates quantum corrections. These terms
are given by 
\bea
\label{eq:a_0}
	a_0[S,\Sigma_g] & =	& \frac{1}{2} \fder{S}{\A_\mu} \{\dd^{\A\A}\} \fder{\Sigma_g}{\A_\mu}
							+ \frac{1}{2} \fder{S}{\C} \{\dd^{\C\C}\} \fder{\Sigma_g}{\C},
\\
\label{eq:a_1}
	a_1[\Sigma_g]	& =	& \frac{1}{2} \fder{}{\A_\mu} \{\dd^{\A\A}\} \fder{\Sigma_g}{\A_\mu}
							+ \frac{1}{2} \fder{}{\C} \{\dd^{\C\C}\} \fder{\Sigma_g}{\C},
\eea
where the $\dd$ represent the ERG kernels and the notation
$\{\dd\}$ denotes their covariantization~\cite{ym,ymi}. 

The natural definitions of functional derivatives of
$\SU(N|N)$ matrices are used~\cite{ymii,SU(N|N),aprop}:
\be
\label{eq:dCdef}
{\delta \over {\delta\C}} \equiv { \left(\!{\begin{array}{cc} {\delta
/ {\delta C^1}} & - {\delta / {\delta \bar{D}}} \\ {\delta /
{\delta D}} & - {\delta / {\delta C^2}} \end{array}} \!\!\right)},
\ee
and from~\eq{expandA}~\cite{SU(N|N),aprop}:
\be
\label{eq:dumbdef}
{\delta\over\delta\A_\mu}\equiv
2T_A{\delta\over\delta\A_{A\,\mu}}+{\sigma\over2N}
{\delta\over\delta\A^0_\mu}.
\ee

The wonderful simplicity of~\eqs{a_0}{a_1} arises
from the realization that the fine detail
of the flow equation
(which, as we will see, does not affect universal
quantities anyway) can be buried in the definition of
the covariantization. Nonetheless, for the purpose of
transparently generalizing to QCD, we now discuss
the covariantization in some detail.
The primary ingredient is
the supercovariantization~\cite{aprop} of the kernel $W$,
$\{W\}_\A$. This is defined according to
\bea
\fl
\label{eq:wv}
	\TiedCovKer{u}{W}{\A}{v} = 
	\sum_{m,n=0}^{\infty} \int_{x_1, \cdots, x_n;y_1,\cdots, y_m;x,y}
	W_{\mu_1\cdots\mu_n,\nu_1\cdots\nu_m}(x_1,\ldots,x_n;y_1,\ldots,y_m;x,y) 
& &
\fl
\nonumber 
\\
	\str \left[u(x)\A_{\mu_1}(x_1)\cdots \A_{\mu_n}(x_n)v(y)\A_{\nu_1}(y_1)\cdots \A_{\nu_m}(y_m) \right],
&&
\eea
where $u$ and $v$ are supermatrix
representations transforming homogeneously as in~\eq{Cgauged} and
where, without loss of generality, we may insist that
$\CovKer{W}{\A}$ satisfies $\TiedCovKer{u}{W}{\A}{v} \equiv
\TiedCovKer{v}{W}{\A}{u}$. For simplicity's sake, we have chosen~\eq{wv}
to contain only a single supertrace. (In the diagrammatic
form of the flow equation, such details
make no difference.)
The $m=n=0$ term is just the original kernel, \ie
\be
\label{eq:mno}
W_,(;;x,y)\equiv W_{xy}.
\ee
The requirement that~\eq{wv} is supergauge invariant enforces a
set of Ward identities on the vertices
$W_{\mu_1\cdots\mu_n,\nu_1\cdots\nu_m}$ which we describe
later. 
The no-$\A^0$ symmetry is obeyed by requiring the coincident line
identities~\cite{ymi}.
These identities are equivalent to the requirement that the gauge
fields all act by commutation~\cite{ymii}, ensuring that the no-$\A^0$
part of~\eq{Agauged} is satisfied. A consequence of
the coincident line identities, which also trivially follows from the
representation of~\eq{wv} in terms of commutators, is that if
$v(y)=\one g(y)$ for all $y$, \ie\ is in the scalar representation
of the gauge group, then the covariantization collapses to
\be
\label{eq:Acoline}
\TiedCovKer{u}{W}{\A}{v}
= (\str u)\ker{W}g,
\ee
where we define
\[
	f \ker{W} g = \int_{x,y} f(x)\, W_{xy}\,g(y) = 
		\int_x f(x) W(-\partial^2/\Lambda^2)\, g(x)
\]
which holds for any momentum space kernel $W(p^2/\Lambda^2)$ and
functions of spacetime $f$, $g$, using
\[
	W_{x y} = W(-\partial^2/\Lambda^2)\,\delta(x-y) 
	= \inte {d^D p \over (2\pi)^D} \, W(p^2/\Lambda^2) \, {\rm e}^{i p \cdot (x-y)}.
\]

At this point, it is instructive to recall
the demonstration of
the $\SU(N|N)$ invariance of the flow equation,
assuming that the covariantizations $\{\dd\}$
are just of the type~\eq{wv}~\cite{aprop}.

Under~\eq{Cgauged}, the $\C$ functional derivative
transforms homogeneously:
\be
\label{eq:dCgauged}
\delta \left({\delta\over\delta\C}\right) =
-i\left[{\delta\over\delta\C},\Omega\right],
\ee
and thus by~\eq{wv}, the corresponding terms 
in~\eqs{a_0}{a_1} are invariant. The $\A$ functional derivative, however,
transforms as~\cite{aprop}:
\be
\label{eq:dAgauged}
\delta \left({\delta\over\delta\A_\mu}\right) =
-i\left[{\delta\over\delta\A_\mu},\Omega\right]
+{i\one\over2N}\tr\left[{\delta\over\delta\A_\mu},\Omega\right].
\ee
The correction is there because~\eq{dumbdef} is traceless, which
in turn is a consequence of the supertracelessness of~\eq{defA}.
The fact that $\delta/\delta\A$ does not transform homogeneously
means that supergauge invariance is destroyed unless the
correction term vanishes for other reasons.

Here, no-$\A^0$ symmetry comes to the rescue. Using the invariance
of~\eq{wv} for homogeneously transforming $u$ and $v$, and the
invariance of $S$ and $\hS$, we have by~\eq{dAgauged} 
and~\eq{Acoline}, that the $\A$ term in~\eq{a_0} transforms to
\be
\label{eq:tlgaugetr}
\delta\left(\frac{\delta S}{\delta {\cal
A}_{\mu}}\{\dd^{\!\A\A}\}\frac{\delta \Sigma_g}{\delta {\cal
A}_{\mu}}\right) = {i\over2N}\, \tr\!\left[{\delta
S\over\delta\A_\mu},\Omega\right]\!\cdot
\dd^{\!\A\A}\!\cdot\str{\delta\Sigma_g\over\delta\A_\mu} +
(S\leftrightarrow\Sigma_g),
\ee
where $S\leftrightarrow\Sigma_g$ stands for the same term with $S$
and $\Sigma_g$ interchanged. But by~\eq{dumbdef} and no-$\A^0$
symmetry,
\[
\str{\delta\Sigma_g\over\delta\A_\mu}={\delta\Sigma_g\over\delta\A^0_\mu}=0
\]
(similarly for $S$), and thus the tree level terms are invariant
under~\eqs{Agauged}{Cgauged}. Likewise, the quantum terms 
in~\eq{a_1} are invariant and this completes the proof
that, for covariantizations of the form~\eq{wv}, the
flow equation is both supergauge and no-$\A^0$ invariant.

As it stands, the covariantization~\eq{wv} is not
general enough for our purposes: we require
the broken phase fields to come with their own
kernels. The first part of the 
solution to this~\cite{aprop} is to
define a new covariantization
\be
\label{eq:wev}
\TiedCovKer{u}{W}{\A\C}{v} = \TiedCovKer{u}{W}{\A}{v}
-{1\over4} \TiedCovKer{[\C,u]}{W_{m}}{\A}{[\C,v]}.
\ee
The $\C$ commutator terms are introduced to allow a difference
between $A$ and $B$ kernels, and $C$ and $D$ kernels, in the
broken phase. They do this because at the level of two-point flow
equations
$\C$ is replaced by $\sigma$ in~\eq{wev}, and $\sigma$
(anti)commutes with the (fermionic) bosonic elements of the
algebra. Thus, extracting the broken phase two-point, classical
flow equations from~\eq{a_0}, we find that the $A^i$ kernels are
given by $\dd^{\!\A\A}$, the $C^i$ kernels by $\dd^{\C\C}$, but
the $B$ kernel is $\dd^{\!\A\A}+\dd^{\!\A\A}_m$ and the $D$ kernel
is $\dd^{\C\C}+\dd^{\C\C}_m$~\cite{aprop}. The $B$ and $D$ kernels can be
combined (\cf~\eqs{F}{Fbar}):
\[
	\dd^{\,F\; \bar{F}}_{MN}(p) =
	\left(
		\begin{array}{cc}
			\dd_p^{B \bar{B}} \delta_{\mu \nu}	& 0
		\\
									0			& -\dd^{D\bar{D}}_p
		\end{array}
	\right).
\]

As must be the case, the extra term in~\eq{wev} is consistent
with both supergauge and no-$\A^0$ invariance.
If $u$ and $v$ transform homogeneously, then so do $[\C,u]$
and $[\C,v]$. Correction terms proportional to the identity 
(\cf~\eq{dAgauged}) are killed by the commutator structure;
this structure also ensures no-$\A^0$ invariance. Notice
that~\eq{Acoline} holds for the extended covariantization.

We are still not quite done: it is convenient to generalize the
covariantization yet further. In particular, since the physical
coupling, $g$, and the unphysical coupling, $g_2$, renormalize
differently~\cite{aprop,mgierg1}, it is useful to furnish
$A^1$ and $A^2$ with different kernels (there is
no need to do this for $C^1$ and $C^2$~\cite{mgierg1}). 
To construct a term which does this, we cannot use
a commutator, since $[\sigma,\delta/\delta A^i]=0$.
 The simplest solution is to
define in~\eqs{a_0}{a_1}
\be
\label{eq:weev}
	u\{\dd^{\A\A}\}v \equiv \TiedCovKer{u}{\dd^{\A\A}}{\A\C}{v} 
		+ \TiedCovKer{u}{\dd^{\A\A}_\sigma}{\A}{\P(v)} 
		+ \TiedCovKer{\P(u)}{\dd^{\A\A}_\sigma}{\A}{v},
\ee
where
\be
\label{eq:Pdef}
8N\,\P(X) = \{\C,X\}\,\str \C - 2\, \C\,\str \C X.
\ee
$\P(X)$ has the following properties
which ensure
both no-$\A^0$ invariance and supergauge invariance:
\numparts
\bea
\label{eq:straceless}
	\str \P(X)	& =	& 0,
\\
\label{eq:oneless}
	\P(\one)	& =	& 0.
\eea
\endnumparts
In the broken phase,
we see that
\[
	\P\left(\fder{}{\A_\mu}\right) = \frac{1}{2} \sigma \fder{}{\A_\mu} + \cdots,
\]
where we have noted from~\eq{dumbdef}
that
$\str \sigma \delta /\delta \A_\mu = 0$
and the ellipsis includes terms with additional fields.
Thus, a minus sign is introduced in the
$A^2$ sector, compared to the $A^1$ sector,
and we find that
$\dd^{A^1A^1} = \dd^{AA} + \dd^{AA}_\sigma$,
$\dd^{A^2A^2} = \dd^{AA} - \dd^{AA}_\sigma$~\cite{mgierg1}.

We conclude this section by commenting on the
Ward identities satisfied by the vertices
of the covariantized kernels. Throughout
this section, we have used the fact that
the $u$ and $v$ in \eg~\eq{weev} are
functional derivatives \wrt, say, $Z^1$
and $Z^2$ to label the kernels \viz\ $\dd^{Z^1Z^2}$.
The diagrammatic form of the Ward identities~\eq{WID-A}
holds for the vertices of the kernels, also,
so long as two of the target fields are identified
with the ends of the kernels \ie\ with $Z^1$
and $Z^2$~\cite{Thesis,mgierg1}.

\subsubsection{Diagrammatics}
\label{sec:YM-Diags}

As mentioned in the introduction, the most
useful representation of the flow equation
is a diagrammatic one, which is shown in
\fig{flow}~\cite{mgierg1,Thesis,Primer,mgiuc}.
\bcf[h]
	\[
	-\flow 
	\dec{
		\ensuremath{\begin{array}{c}\begin{picture}(0,0)%
\includegraphics{pstex/Vertex-S.pstex}%
\end{picture}%
\setlength{\unitlength}{3947sp}%
\begingroup\makeatletter\ifx\SetFigFont\undefined%
\gdef\SetFigFont#1#2#3#4#5{%
  \reset@font\fontsize{#1}{#2pt}%
  \fontfamily{#3}\fontseries{#4}\fontshape{#5}%
  \selectfont}%
\fi\endgroup%
\begin{picture}(341,318)(2180,-963)
\put(2291,-859){\makebox(0,0)[lb]{\smash{{\SetFigFont{11}{13.2}{\rmdefault}{\mddefault}{\updefault}{\color[rgb]{0,0,0}$S$}%
}}}}
\end{picture}%
 \end{array}}
	}{\{f\}} =
	\frac{1}{2}
	\dec{
		\ensuremath{\begin{array}{c}\input{pstex/Dumbbell-S-Sigma_g.pstex_t} \end{array}} - \ensuremath{\begin{array}{c}\input{pstex/Padlock-Sigma_g.pstex_t} \end{array}}
	}{\{f\}}
	\]
\caption{The diagrammatic form of the flow equation.}
\label{fig:flow}
\ecf

The term on the \lhs\ generates the flow of all cyclically independent
Wilsonian effective action
vertex coefficient functions which correspond
to the set of broken phase fields $\{f\}$.

The objects on the \rhs\ of \fig{flow}
have two different types of component. The lobes
represent vertices of action functionals.
The object attaching
to the various lobes, \DummyKernel,  is
the sum over vertices of the covariantized 
ERG kernels~\cite{ymi,aprop}
and, like the action vertices, can be 
decorated by fields belonging to $\{f\}$.
The fields of the action vertex (vertices) to
which the vertices of the kernels attach
act as labels for the ERG kernels.
We henceforth 
loosely refer to both individual and summed over 
vertices of the kernels simply as a kernel. 
The dumbbell-like term corresponds
to the classical term, $a_0$, whereas the padlock-like
diagram corresponds to the quantum term, $a_1$.%
\footnote{There is an additional, improperly regularized
term generated by the flow equation which has
been removed by a suitable constraint on the 
covariantization~\cite{ym,aprop,mgierg1,mgiuc}.}
The rule for decorating the classical and quantum terms 
is simple: the set of fields, $\{f\}$, are distributed in 
all independent ways between the component objects of each diagram.

Embedded within the diagrammatic rules is a prescription for evaluating the
group theory factors. 
Suppose that we wish to focus on the flow of a particular
vertex coefficient function which, necessarily, has a unique
supertrace structure. 
For example, we might be interested in just
the $S^{C^1 C^1}$ component of~\eq{Vertex-C1C1}.
On the \rhs\ of the flow equation, we must
focus on the components of each diagram
with precisely the same
supertrace structure as the \lhs,
noting that the kernel, like the vertices,
has multi-supertrace contributions.
In this more explicit diagrammatic picture,
the kernel is to be considered a double
sided object (for more
details see~\cite{Thesis,mgierg1}).
Thus, whilst the dumbbell like term of \fig{flow}
has at least one associated supertrace, the next diagram
has  at least two, on account of the loop
(this is strictly true only in the
case that kernel attaches to fields on the same
supertrace). If a closed
circuit formed by a kernel is devoid
of fields then it contributes
a group theory factor, depending on
the flavours of the fields to which the kernel forming
the loop attaches. This is most easily appreciated by
noting that $\str \sigma_\pm = \pm N$ (see~\eq{sigma_pm}).
In the counterclockwise sense, a $\sigma_+$
can always be inserted for free after an $A^1$, $C^1$ or $\bar{F}$,
whereas a $\sigma_-$
can always be inserted for free after an $A^2$, $C^2$ or $F$. 

The above prescription for evaluating the
group theory factors receives $1/N$ corrections in 
the $A^1$ and $A^2$ sectors, as a consequence of
the $\SU(N)$ completeness relation~\cite{ymi}. If a kernel
attaches to an $A^1$ or $A^2$, it comprises a direct
attachment and an indirect attachment.
In the former case, one supertrace associated
with some vertex coefficient function  is `broken
open' by an end of a kernel: the fields on
this supertrace and the single supertrace component of the
kernel are on the same circuit.
In the latter case, the kernel does not break anything open
and so the two sides of the kernel pinch together
at the end associated with the indirect attachment.
This is illustrated
in \fig{Attach}; for more detail, see~\cite{Thesis,mgierg1,evalues}.
\bcf[h]
	\[
		\ensuremath{\begin{array}{c}\begin{picture}(0,0)%
\includegraphics{pstex/Direct.pstex}%
\end{picture}%
\setlength{\unitlength}{3947sp}%
\begingroup\makeatletter\ifx\SetFigFont\undefined%
\gdef\SetFigFont#1#2#3#4#5{%
  \reset@font\fontsize{#1}{#2pt}%
  \fontfamily{#3}\fontseries{#4}\fontshape{#5}%
  \selectfont}%
\fi\endgroup%
\begin{picture}(624,477)(2089,-976)
\end{picture}
 \end{array}} \rightarrow \left.\ensuremath{\begin{array}{c}\begin{picture}(0,0)%
\includegraphics{pstex/Direct.pstex}%
\end{picture}%
\setlength{\unitlength}{3947sp}%
\begingroup\makeatletter\ifx\SetFigFont\undefined%
\gdef\SetFigFont#1#2#3#4#5{%
  \reset@font\fontsize{#1}{#2pt}%
  \fontfamily{#3}\fontseries{#4}\fontshape{#5}%
  \selectfont}%
\fi\endgroup%
\begin{picture}(624,477)(2089,-976)
\end{picture}
 \end{array}}\right|_{\mathrm{direct}} + \frac{1}{N} \left[ \ensuremath{\begin{array}{c}\begin{picture}(0,0)%
\includegraphics{pstex/Indirect-2.pstex}%
\end{picture}%
\setlength{\unitlength}{3947sp}%
\begingroup\makeatletter\ifx\SetFigFont\undefined%
\gdef\SetFigFont#1#2#3#4#5{%
  \reset@font\fontsize{#1}{#2pt}%
  \fontfamily{#3}\fontseries{#4}\fontshape{#5}%
  \selectfont}%
\fi\endgroup%
\begin{picture}(624,809)(2089,-1279)
\put(2355,-554){\makebox(0,0)[lb]{\smash{\SetFigFont{8}{9.6}{\rmdefault}{\mddefault}{\updefault}{\color[rgb]{0,0,0}$A^2$}%
}}}
\end{picture}
 \end{array}} - \ensuremath{\begin{array}{c}\begin{picture}(0,0)%
\includegraphics{pstex/Indirect-1.pstex}%
\end{picture}%
\setlength{\unitlength}{3947sp}%
\begingroup\makeatletter\ifx\SetFigFont\undefined%
\gdef\SetFigFont#1#2#3#4#5{%
  \reset@font\fontsize{#1}{#2pt}%
  \fontfamily{#3}\fontseries{#4}\fontshape{#5}%
  \selectfont}%
\fi\endgroup%
\begin{picture}(624,809)(2089,-1279)
\put(2355,-554){\makebox(0,0)[lb]{\smash{\SetFigFont{8}{9.6}{\rmdefault}{\mddefault}{\updefault}{\color[rgb]{0,0,0}$A^1$}%
}}}
\end{picture}
 \end{array}} \right]
	\]
\caption{The $1/N$ corrections to the group theory factors.}
\label{fig:Attach}
\ecf 

We can thus consider the diagram on the \lhs\ as having been unpackaged,
to give the terms on the \rhs. The dotted lines in the diagrams with indirect
attachments serve to remind us where the loose end of the kernel attaches
in the parent diagram.

\subsection{Adding Quarks}

The game now is easy. We add to the flow equation
classical and quantum terms for the fields $\Up$, $\Down$
and modify the superhiggs sector. For all fields, we  ensure 
that there is sufficient
freedom in the covariantization to allow enough different
kernels for the broken phase fields.

Though the details will ultimately be hidden
in the diagrammatic form of the flow equation,
we will give examples of choices we can make
for the covariantization. Just as in the pure
Yang-Mills case, the kernels of the propagating
fields may turn out to 
be linear combinations of those
which appear in the flow equation. 
Though we will not explicitly perform this
change of basis, it is instructive to see
how we can, if we so desire, construct
the covariantization so as to make this procedure
as easy as possible. 

As in the pure Yang-Mills case, the starting
point for constructing the covariantization 
is~\eq{wv}. To this we now add additional
terms which reflect the complete breaking
of the unphysical $\SU(N)$ gauge symmetry.
Knowing that $B_\mu$ decomposes into $N$ flavours,
it makes sense to add
to the covariantization a term of the form
\[
	-\frac{1}{4}
	\sum_{j=1}^N 
	\left( 
		\TiedCovKer{[\P_j,[\C,u]]}{\dd^{\A\A}_j}{\A}{[\P_j,[\C,v]]} 
	\right),
\]
where we have used~\eq{Projectors}.
(As usual, the overall factor is merely a matter of convention.)
The presence of the $\C$ commutators is purely
for convenience. In the broken phase, they
project on to the block off-diagonal components
of $u$ and $v$, which ensures that, at the two-point
level, the above term
does not interfere with the flow of
of the components of the field $A_\mu^2$. This
makes it easier to extract the kernels of
the propagating fields in terms of the kernels
in the flow equation.

The gauge field $A^2_\mu$ has $N^2-1$
independent components. Since we can use
the same kernel for a field and its
Hermitian conjugate, we require a total
of $N(N+1)/2-1$ kernels.
So, for the components of $A_\mu^2$, we 
add to the covariantization a term of
the form:
\be
\label{eq:weeev}
	\sum_{j=1}^{N(N+1)/2-1}
	\left[
		\TiedCovKer{u}{\dd^{\A\A}_{j+N}}{\A}{\P'_j(\P(v))} 
		+ \TiedCovKer{\P'_j(\P(u))}{\dd^{\A\A}_{j+N}}{\A}{v}
	\right],
\ee
where
\[
	\P'_j(X) = (Y_jXZ_j + Z_jXY_j)\str Y_jZ_j - Y_jZ_j\str Y_jXZ_j -Z_jY_j \str Z_jXY_j,
\]
and the 
$Y_j$ and $Z_j$ contain linear combinations 
of the symmetric phase versions
of the projectors defined by~\eq{Projectors}
in their bottom right block (all other elements
being zero).
There are clearly many different choices we
can take for $Y_j$ and $Z_j$. An example for $N>2$
would be to set $Y_j = Z_j$ and choose 
the first $N(N-1)/2$
$Y_j$ to be such that, in the the broken
phase, they reduce to 
the $\nCr{N}{2}$ independent
combinations of projectors of
the form $Y_j = P_k + P_{l\neq k}$.
Then we can take the remaining $Y_j$ to
be any $N-1$ of the $N$ $P_i$.
Notice that $\P'$ satisfies the 
conditions~\eqs{straceless}{oneless} and thus does
not spoil either the supergauge or no-$\A^0$
invariance of the flow equation.
The appearance of $\P$ (see~\eq{Pdef})
in~\eq{weeev} is again for convenience ensuring that,
at the two-point level, the kernels of~\eq{weeev} do not
appear in the flow of the components of $B_\mu$.

Finally, then, a suitable choice for the covariantization 
of $\dd^{\A\A}$ is:
\beas
\lefteqn{
	u\{\dd^{\A\A}\}v \equiv \TiedCovKer{u}{\dd^{\A\A}}{\A}{v} 	
	-\frac{1}{4}
	\sum_{j=1}^N 
	\left( 
		\TiedCovKer{[\P_j,[\C,u]]}{\dd^{\A\A}_j}{\A}{[\P_j,[\C,v]]} 
	\right)
}
\\
&&
		\qquad \qquad \qquad
		+
		\sum_{j=1}^{N(N+1)/2-1}
		\left[
			\TiedCovKer{u}{\dd^{\A\A}_{j+N}}{\A}{\P'_j(\P(v))} 
			+ \TiedCovKer{\P'_j(\P(u))}{\dd^{\A\A}_{j+N}}{\A}{v}
		\right].
\eeas

The modifications to the covariantization in the superhiggs
sector
are almost identical; the only real difference is that,
since the superscalars are not supertraceless, there are more propagating
degrees of freedom in than in the $\A_\mu$ sector, and so
we must introduce additional kernels to take account of this.

The inclusion of quarks follows a similar pattern. 
The contributions of the up-like quarks to the flow equation
are the standard ones for spinor fields~\cite{giqed}%
\footnote{
Though, compared to QED, there is no need to 
explicitly take care of the anticommuting nature of
the quark fields, since
this is automatically taken care of by the embedding
into the supergroup.
}, 
with the contribution
to the classical term given by
\[
	\frac{1}{2}
	\left(
		 \fder{S}{\antiUp} \{\dd^{\antiUp \Up} \} \fder{\Sigma_g}{\Up}
		+
		\fder{S}{\Up} \{\dd^{\antiUp \Up} \} \fder{\Sigma_g}{\antiUp}
	\right),
\]
and  the contribution to quantum term given by
\[
	\frac{1}{2}
	\left(
		\fder{}{\Up} \{\dd^{\antiUp \Up} \} \fder{\Sigma_g}{\antiUp}
		+ 
		\fder{}{\antiUp} \{\dd^{\antiUp \Up} \} \fder{\Sigma_g}{\Up}
	\right),
\]
where we have suppressed spinor indices and functional derivatives
\wrt\ $\Psi$ are defined as for any other
unconstrained (\ie\ not supertraceless) superfield (see~\eq{dCdef}).

The covariantization is chosen
to be:
\be
\fl
	u \{\dd^{\antiUp \Up}\} v = 
		\TiedCovKer{u}{\dd^{\antiUp \Up}}{\A}{v}
	+ \sum_{j=1}^{N} \TiedCovKer{[\P_j,\varpi_{-}(u)]}{\dd^{\antiUp \Up}_{j}}{\A}{[\P_j,\varpi_{+} (v)]}
	+\ldots
\label{eq:QuarkCov}
\ee
The first term on the \rhs\ is the usual contribution involving just
the supercovariantization. In the second term we have introduced,
for convenience, the objects
\[
	\varpi_{\pm}(X) = \frac{1}{8} 
	\left(
		[\C,[\C,X]] \pm 2[\C,X]]
	\right)
\]
defined so that, in the broken phase, they reduce to $\sigma_+ X \sigma_-$
and $\sigma_- X\sigma_+$ (plus interaction terms). 
Consequently, at the two-point level,
the second term on the \rhs\ of~\eq{QuarkCov}
contributes only to the flow of the physical
quarks. 
The ellipsis represents additional terms
which provide kernels for the rest of the propagating
fields embedded in $\Up$. 

\subsection{Diagrammatics for QCD}

\subsubsection{The Exact Flow Equation}

The beauty of the diagrammatic form of the flow equation
given in \fig{flow} is that it can be directly
generalized from $\SU(N)$ Yang-Mills to QCD:
all we need to do is to extend
the set of broken phase fields which 
can decorate the diagrams
both as internal and external fields. A consequence
of this is that the prescription for extracting
the group theory factors receives further
corrections, which follow from inserting
the appropriate projectors to go from derivatives
\wrt\ supermatrix block fields to derivatives
\wrt\ their appropriate components.

In fact, it is useful to employ a slightly
different flow equation, in which we work directly
with renormalized fields~\cite{scalar1,scalar2,giqed} 
(recall that the components of all fields bar 
$\Up$ and $\Down$ are protected from field
strength renormalization).
This flow equation is
given in~\fig{flowQCD}. It is not the result of
scaling the wavefunction renormalizations out
of the version of~\eq{flow}
appropriate to QCD but is a perfectly valid
flow equation, nonetheless~\cite{scalar1,scalar2}. 
This is a manifestation
of the tremendous freedom 
we have in constructing flow equations,
encapsulated in~\eq{blocked}.
\bcf[h]
	\[
	\ds
	\left(
		-\flow + \sum_{\chi \epsilon \{f\}} \gamma^{(\chi)}
	\right) 
	\dec{
		\ensuremath{\begin{array}{c}\begin{picture}(0,0)%
\includegraphics{pstex/Vertex-S.pstex}%
\end{picture}%
\setlength{\unitlength}{3947sp}%
\begingroup\makeatletter\ifx\SetFigFont\undefined%
\gdef\SetFigFont#1#2#3#4#5{%
  \reset@font\fontsize{#1}{#2pt}%
  \fontfamily{#3}\fontseries{#4}\fontshape{#5}%
  \selectfont}%
\fi\endgroup%
\begin{picture}(341,318)(2180,-963)
\put(2291,-859){\makebox(0,0)[lb]{\smash{{\SetFigFont{11}{13.2}{\rmdefault}{\mddefault}{\updefault}{\color[rgb]{0,0,0}$S$}%
}}}}
\end{picture}%
 \end{array}}
	}{\{f\}} =
\\
	\ds
	\frac{1}{2}
	\dec{
		\ensuremath{\begin{array}{c}\input{pstex/Dumbbell-S-Sigma_g.pstex_t} \end{array}} - \ensuremath{\begin{array}{c}\input{pstex/Padlock-Sigma_g.pstex_t} \end{array}}
	}{\{f\}}
	\]
\caption{The diagrammatic form of the flow equation.}
\label{fig:flowQCD}
\ecf

The term $\sum_{\chi \epsilon \{f\}} \gamma^{(\chi)}$
explicitly takes account of the anomalous
dimensions of the fields which suffer
field strength renormalization.
The field $\chi$ belongs to the set of fields 
$\{f\}$ and
the notation $\gamma^{(\chi)}$
just stands for the anomalous dimension of
the field $\chi$ (which is zero for all
but the components of $\Quarks_{u,d}$).

\subsubsection{Perturbative Diagrammatics}

In the perturbative domain, we have the following
weak coupling expansions~\cite{ymi,aprop,Thesis,mgierg1}.
The Wilsonian effective action is given by
\be
	S = \sum_{i=0}^\infty g^{2(i-1)} S_i = \frac{1}{g^2}S_0 + S_1 + \cdots,
\label{eq:Weak-S}
\ee
where $S_0$ is the classical effective action and the $S_{i>0}$
the $i$th-loop corrections. The seed action has a similar expansion:
\be
	\hat{S} = \sum_{i=0}^\infty  g^{2i}\hat{S}_i,
\label{eq:Weak-hS}
\ee
and the $\beta$ function and anomalous dimensions are defined as usual:
\bea
	\beta & \equiv & \flow g = \sum_{i=1}^\infty  g^{2i+1} \beta_i,
\label{eq:beta}
\\[1ex]
	\gamma^{(\chi)} & \equiv & \flow \ln Z^{(\chi)} =\sum_{i=1}^{\infty}  g^{2i} \gamma_i^{(\chi)},
\label{eq:AD}
\eea
where $Z^{(\chi)}$ is the field strength renormalization
of the field of species $\chi$.

We also introduce $\beta$ functions for the dimensionless
mass parameters, $\overline{m}^i \equiv m^i / \Lambda$:
\be
	\beta^j \equiv \flow \overline{m}^j 	= \sum_{i=1}^{\infty}  g^{2i} \beta_i^j.
\label{eq:gamma}
\ee

Defining $\Sigma_i = S_i - 2\hS_i$, the weak coupling flow equations
follow from substituting~\eq{Weak-S}--\eq{gamma}
into the flow equation, as shown in 
\fig{WeakCouplingFE}.

\bcf[h]
	\[
	\fl
		\dec{
			\ensuremath{\begin{array}{c}\input{pstex/Vertex-n-LdL.pstex_t} \end{array}} 
		}{\{f\}}
		= 
		\dec{
			\begin{array}{c}
				\ds
				\sum_{r=1}^n 
				\left[
					2\left(n_r -1 \right) \beta_r 
					+\sum_j \beta_r^j \pder{}{\overline{m}^j} 
					-\sum_{\chi \epsilon \{f\}} \gamma^{(\chi)}_r
				\right]\ensuremath{\begin{array}{c}\begin{picture}(0,0)%
\includegraphics{pstex/Vertex-n_r-B.pstex}%
\end{picture}%
\setlength{\unitlength}{3947sp}%
\begingroup\makeatletter\ifx\SetFigFont\undefined%
\gdef\SetFigFont#1#2#3#4#5{%
  \reset@font\fontsize{#1}{#2pt}%
  \fontfamily{#3}\fontseries{#4}\fontshape{#5}%
  \selectfont}%
\fi\endgroup%
\begin{picture}(320,318)(1776,-676)
\put(1857,-545){\makebox(0,0)[lb]{\smash{\SetFigFont{11}{13.2}{\rmdefault}{\mddefault}{\updefault}{\color[rgb]{0,0,0}$n_r$}%
}}}
\end{picture}
 \end{array}} 
			\\[4ex]
				\ds
				+ \frac{1}{2} 
				\left( 
					\sum_{r=0}^n \ensuremath{\begin{array}{c}\input{pstex/Dumbbell-n_r-r.pstex_t} \end{array}} - \ensuremath{\begin{array}{c}\input{pstex/Vertex-Sigma_n_-B.pstex_t} \end{array}}
				\right)
			\end{array}
		}{\{f\}}
	\]
\caption{The weak coupling flow equations.}
\label{fig:WeakCouplingFE}
\ecf

The symbol $\bullet$ means
$-\flowConstAlj$. We will see shortly why the
notation for the ERG kernels, \DummyKernel,
includes this symbol.
A vertex whose argument is an unadorned letter, say $n$,
represents $S_n$. We define $n_r \equiv n-r$ and $n_\pm \equiv n \pm 1$. The
bar notation of the dumbbell term is defined as follows:
\[
	a_0[\bar{S}_{n-r}, \bar{S}_r] 	\equiv 	a_0[S_{n-r}, S_r] - a_0[S_{n-r}, \hat{S}_r] - a_0[\hat{S}_{n-r}, S_r].
\]

We illustrate the use of the flow equation
by considering the flow of all two-point, classical vertices.
This is done by setting $n=0$ in~\fig{WeakCouplingFE}
and specializing $\{f\}$ to contain two fields, 
as shown in \fig{TLTPs}.
We note that we can and do choose
all such vertices to be single supertrace terms~\cite{Thesis,mgierg1}.
\bcf[h]
	\be
		\ensuremath{\begin{array}{c}\input{pstex/Vertex-TLTP-LdL.pstex_t} \end{array}} = \ensuremath{\begin{array}{c}\input{pstex/Dumbbell-S_0-S_0.pstex_t} \end{array}} - \ensuremath{\begin{array}{c}\input{pstex/Dumbbell-S_0-hS_0.pstex_t} \end{array}} - \ensuremath{\begin{array}{c}\input{pstex/Dumbbell-hS_0-S_0.pstex_t} \end{array}}
	\label{eq:TLTP-flow}
	\ee
\caption{Flow of all possible two-point, classical vertices.}
\label{fig:TLTPs}
\ecf

Following~\cite{ym,ymi,ymii,aprop,Thesis,mgierg1,scalar2}, 
we use the freedom inherent in $\hat{S}$ by choosing the two-point, classical
seed action vertices equal to the corresponding Wilsonian effective
action vertices. Equation~\eq{TLTP-flow} now simplifies.
Rearranging, integrating \wrt\ $\Lambda$ and choosing the appropriate
integration constants~\cite{Thesis,mgierg1}, we arrive at the following
relationship between the integrated ERG kernels and the two-point,
classical vertices. 
\be
		\ensuremath{\begin{array}{c}\input{pstex/EffPropReln.pstex_t} \end{array}}	= \ensuremath{\begin{array}{c}\begin{picture}(0,0)%
\includegraphics{pstex/K-Delta.pstex}%
\end{picture}%
\setlength{\unitlength}{3947sp}%
\begingroup\makeatletter\ifx\SetFigFont\undefined%
\gdef\SetFigFont#1#2#3#4#5{%
  \reset@font\fontsize{#1}{#2pt}%
  \fontfamily{#3}\fontseries{#4}\fontshape{#5}%
  \selectfont}%
\fi\endgroup%
\begin{picture}(374,395)(1791,-1006)
\put(1791,-843){\makebox(0,0)[lb]{\smash{{\SetFigFont{8}{9.6}{\rmdefault}{\mddefault}{\updefault}{\color[rgb]{0,0,0}$M$}%
}}}}
\end{picture}%
 \end{array}} - \ensuremath{\begin{array}{c}\begin{picture}(0,0)%
\includegraphics{pstex/FullGaugeRemainder.pstex}%
\end{picture}%
\setlength{\unitlength}{3947sp}%
\begingroup\makeatletter\ifx\SetFigFont\undefined%
\gdef\SetFigFont#1#2#3#4#5{%
  \reset@font\fontsize{#1}{#2pt}%
  \fontfamily{#3}\fontseries{#4}\fontshape{#5}%
  \selectfont}%
\fi\endgroup%
\begin{picture}(424,395)(2053,-930)
\put(2053,-773){\makebox(0,0)[lb]{\smash{{\SetFigFont{8}{9.6}{\rmdefault}{\mddefault}{\updefault}{\color[rgb]{0,0,0}$M$}%
}}}}
\end{picture}%
 \end{array}}
							= \ensuremath{\begin{array}{c}\begin{picture}(0,0)%
\includegraphics{pstex/K-Delta.pstex}%
\end{picture}%
\setlength{\unitlength}{3947sp}%
\begingroup\makeatletter\ifx\SetFigFont\undefined%
\gdef\SetFigFont#1#2#3#4#5{%
  \reset@font\fontsize{#1}{#2pt}%
  \fontfamily{#3}\fontseries{#4}\fontshape{#5}%
  \selectfont}%
\fi\endgroup%
\begin{picture}(374,395)(1791,-1006)
\put(1791,-843){\makebox(0,0)[lb]{\smash{{\SetFigFont{8}{9.6}{\rmdefault}{\mddefault}{\updefault}{\color[rgb]{0,0,0}$M$}%
}}}}
\end{picture}%
 \end{array}} - \ensuremath{\begin{array}{c}\begin{picture}(0,0)%
\includegraphics{pstex/DecomposedGR.pstex}%
\end{picture}%
\setlength{\unitlength}{3947sp}%
\begingroup\makeatletter\ifx\SetFigFont\undefined%
\gdef\SetFigFont#1#2#3#4#5{%
  \reset@font\fontsize{#1}{#2pt}%
  \fontfamily{#3}\fontseries{#4}\fontshape{#5}%
  \selectfont}%
\fi\endgroup%
\begin{picture}(540,395)(1936,-925)
\put(1936,-776){\makebox(0,0)[lb]{\smash{{\SetFigFont{8}{9.6}{\rmdefault}{\mddefault}{\updefault}{\color[rgb]{0,0,0}$M$}%
}}}}
\end{picture}%
 \end{array}}
	\label{eq:EPReln-A}
\ee
We have attached the integrated ERG kernel,
denoted by a solid line, 
to an arbitrary structure since it only
ever appears as an internal line.
The field labelled by $M$ can be any of the broken phase
fields. The object $\GR \!\! \equiv \; \GRkpr \!\!\! \GRk$ 
is a gauge remainder (\cf~\eq{A1-GR}).
The gauge remainder components are 
non-null only in the sectors corresponding
to (components of) $A^i_\mu$ and $F_R$
and, in these sectors, $\GRk$ and
$\GRkpr$ are related as a consequence of
gauge invariance, as we will see shortly. Note that,
in the case that a full gauge remainder bites a vertex,
as opposed to just a $\GRk$, we can replace the half
arrows on the \rhs\ of~\eq{WID-A} (which we
recall just indicate to former presence of a $\GRk$) 
with a $\GRkpr$~\cite{Thesis,mgierg1}.

We have been able to construct
the effective propagator for each and every independent
classical, two-point vertex because we ensured that,
for each such vertex, there exists an independent
(integrated) kernel.

From the effective propagator relation 
and~\eq{D-ID-GR-TP-A} follows a 
series of diagrammatic identities.
In QCD, as opposed to pure Yang-Mills,
the renormalization of $\expectation{C}_{u,d}$---equivalently
the renormalization of the quark masses---means that
one-point $C_{u,d}^{1,2}$ vertices exist beyond
tree level, spoiling~\eq{D-ID-GR-TP-A} at the loop level.
The first of the diagrammatic identities is, then,
the classical part of~\eq{D-ID-GR-TP-A}:
\be
	\ensuremath{\begin{array}{c}\begin{picture}(0,0)%
\includegraphics{pstex/GR-TLTP.pstex}%
\end{picture}%
\setlength{\unitlength}{3947sp}%
\begingroup\makeatletter\ifx\SetFigFont\undefined%
\gdef\SetFigFont#1#2#3#4#5{%
  \reset@font\fontsize{#1}{#2pt}%
  \fontfamily{#3}\fontseries{#4}\fontshape{#5}%
  \selectfont}%
\fi\endgroup%
\begin{picture}(757,318)(1880,-963)
\put(2296,-857){\makebox(0,0)[lb]{\smash{{\SetFigFont{11}{13.2}{\rmdefault}{\mddefault}{\updefault}{\color[rgb]{0,0,0}$0$}%
}}}}
\end{picture}%
 \end{array}} = 0.
\label{eq:GR-TLTP-A}
\ee

From the effective propagator relation and~\eq{GR-TLTP-A},
two further diagrammatic identities follow.
First, consider attaching
an effective propagator to the right-hand field in~\eq{GR-TLTP-A}
and applying
the effective propagator before $\GRk$ has acted. Diagrammatically,
this gives
\[
	\ensuremath{\begin{array}{c}\begin{picture}(0,0)%
\includegraphics{pstex/GR-TLTP-EP.pstex}%
\end{picture}%
\setlength{\unitlength}{3947sp}%
\begingroup\makeatletter\ifx\SetFigFont\undefined%
\gdef\SetFigFont#1#2#3#4#5{%
  \reset@font\fontsize{#1}{#2pt}%
  \fontfamily{#3}\fontseries{#4}\fontshape{#5}%
  \selectfont}%
\fi\endgroup%
\begin{picture}(1081,306)(2490,-1356)
\put(2776,-1250){\makebox(0,0)[lb]{\smash{\SetFigFont{11}{13.2}{\rmdefault}{\mddefault}{\updefault}{\color[rgb]{0,0,0}0}%
}}}
\end{picture}
 \end{array}} = 0 = \ensuremath{\begin{array}{c}\begin{picture}(0,0)%
\includegraphics{pstex/k.pstex}%
\end{picture}%
\setlength{\unitlength}{3947sp}%
\begingroup\makeatletter\ifx\SetFigFont\undefined%
\gdef\SetFigFont#1#2#3#4#5{%
  \reset@font\fontsize{#1}{#2pt}%
  \fontfamily{#3}\fontseries{#4}\fontshape{#5}%
  \selectfont}%
\fi\endgroup%
\begin{picture}(174,174)(2239,-821)
\end{picture}
 \end{array}} - \ensuremath{\begin{array}{c}\begin{picture}(0,0)%
\includegraphics{pstex/kkprk.pstex}%
\end{picture}%
\setlength{\unitlength}{3947sp}%
\begingroup\makeatletter\ifx\SetFigFont\undefined%
\gdef\SetFigFont#1#2#3#4#5{%
  \reset@font\fontsize{#1}{#2pt}%
  \fontfamily{#3}\fontseries{#4}\fontshape{#5}%
  \selectfont}%
\fi\endgroup%
\begin{picture}(499,174)(2239,-820)
\end{picture}
 \end{array}},
\]
which implies the following diagrammatic identity:
\be
	\ensuremath{\begin{array}{c}\begin{picture}(0,0)%
\includegraphics{pstex/GR-relation.pstex}%
\end{picture}%
\setlength{\unitlength}{3947sp}%
\begingroup\makeatletter\ifx\SetFigFont\undefined%
\gdef\SetFigFont#1#2#3#4#5{%
  \reset@font\fontsize{#1}{#2pt}%
  \fontfamily{#3}\fontseries{#4}\fontshape{#5}%
  \selectfont}%
\fi\endgroup%
\begin{picture}(314,174)(2239,-821)
\end{picture}
 \end{array}} = 1.
\label{eq:GR-relation-A}
\ee

The effective propagator relation, together
with~\eq{GR-relation-A}, implies that
\[
	\ensuremath{\begin{array}{c}\begin{picture}(0,0)%
\includegraphics{pstex/TLTP-EP-GR.pstex}%
\end{picture}%
\setlength{\unitlength}{3947sp}%
\begingroup\makeatletter\ifx\SetFigFont\undefined%
\gdef\SetFigFont#1#2#3#4#5{%
  \reset@font\fontsize{#1}{#2pt}%
  \fontfamily{#3}\fontseries{#4}\fontshape{#5}%
  \selectfont}%
\fi\endgroup%
\begin{picture}(1059,306)(2512,-1356)
\put(2776,-1250){\makebox(0,0)[lb]{\smash{\SetFigFont{11}{13.2}{\rmdefault}{\mddefault}{\updefault}{\color[rgb]{0,0,0}0}%
}}}
\end{picture}
 \end{array}} = \ensuremath{\begin{array}{c}\begin{picture}(0,0)%
\includegraphics{pstex/kpr.pstex}%
\end{picture}%
\setlength{\unitlength}{3947sp}%
\begingroup\makeatletter\ifx\SetFigFont\undefined%
\gdef\SetFigFont#1#2#3#4#5{%
  \reset@font\fontsize{#1}{#2pt}%
  \fontfamily{#3}\fontseries{#4}\fontshape{#5}%
  \selectfont}%
\fi\endgroup%
\begin{picture}(174,174)(2379,-821)
\end{picture}
 \end{array}} - \ensuremath{\begin{array}{c}\begin{picture}(0,0)%
\includegraphics{pstex/kprkkpr.pstex}%
\end{picture}%
\setlength{\unitlength}{3947sp}%
\begingroup\makeatletter\ifx\SetFigFont\undefined%
\gdef\SetFigFont#1#2#3#4#5{%
  \reset@font\fontsize{#1}{#2pt}%
  \fontfamily{#3}\fontseries{#4}\fontshape{#5}%
  \selectfont}%
\fi\endgroup%
\begin{picture}(508,174)(2379,-817)
\end{picture}
 \end{array}} = 0.
\]
In other words, the (non-zero) structure $\ensuremath{\begin{array}{c}\begin{picture}(0,0)%
\includegraphics{pstex/EP-GR.pstex}%
\end{picture}%
\setlength{\unitlength}{3947sp}%
\begingroup\makeatletter\ifx\SetFigFont\undefined%
\gdef\SetFigFont#1#2#3#4#5{%
  \reset@font\fontsize{#1}{#2pt}%
  \fontfamily{#3}\fontseries{#4}\fontshape{#5}%
  \selectfont}%
\fi\endgroup%
\begin{picture}(406,118)(3165,-1234)
\end{picture}
 \end{array}}$ kills
a classical, two-point vertex. But, by~\eq{GR-TLTP-A}, 
this suggests that the structure $\ensuremath{\begin{array}{c}\begin{picture}(0,0)%
\includegraphics{pstex/EP-GR.pstex}%
\end{picture}%
\setlength{\unitlength}{3947sp}%
\begingroup\makeatletter\ifx\SetFigFont\undefined%
\gdef\SetFigFont#1#2#3#4#5{%
  \reset@font\fontsize{#1}{#2pt}%
  \fontfamily{#3}\fontseries{#4}\fontshape{#5}%
  \selectfont}%
\fi\endgroup%
\begin{picture}(406,118)(3165,-1234)
\end{picture}
 \end{array}}$
must be equal, up to some factor, to $\lhd$. Hence, 
\be
	\ensuremath{\begin{array}{c}\begin{picture}(0,0)%
\includegraphics{pstex/EP-GRpr.pstex}%
\end{picture}%
\setlength{\unitlength}{3947sp}%
\begingroup\makeatletter\ifx\SetFigFont\undefined%
\gdef\SetFigFont#1#2#3#4#5{%
  \reset@font\fontsize{#1}{#2pt}%
  \fontfamily{#3}\fontseries{#4}\fontshape{#5}%
  \selectfont}%
\fi\endgroup%
\begin{picture}(628,174)(1926,-821)
\end{picture}
 \end{array}} \equiv \ensuremath{\begin{array}{c}\begin{picture}(0,0)%
\includegraphics{pstex/GR-PEP.pstex}%
\end{picture}%
\setlength{\unitlength}{3947sp}%
\begingroup\makeatletter\ifx\SetFigFont\undefined%
\gdef\SetFigFont#1#2#3#4#5{%
  \reset@font\fontsize{#1}{#2pt}%
  \fontfamily{#3}\fontseries{#4}\fontshape{#5}%
  \selectfont}%
\fi\endgroup%
\begin{picture}(786,174)(2089,-821)
\end{picture}
 \end{array}},
\label{eq:PseudoEP-A}
\ee
where the dot-dash line represents the pseudo effective propagators 
of~\cite{Thesis,mgierg1}. 

In practice,
pseudo effective propagators only ever appear in a very specific
way~\cite{mgiuc}, which we now describe. Consider a three-point vertex
attached to two arbitrary structures, $A$ and $B$, by two effective
propagators. The third field on the vertex is taken to be an $A^1$ carrying
momentum $p$ and we suppose
that we now Taylor expand the vertex to zeroth order in $p$. 
Using~\eq{Taylor-A} we have:
\be
\label{eq:EP-dCTP-EP}
	\ensuremath{\begin{array}{c}\input{pstex/A-EP-dCTP-EP-B.pstex_t} \end{array}} = 
	\begin{array}{cccccc}
		-	& \ensuremath{\begin{array}{c}\input{pstex/A-dEP-B.pstex_t} \end{array}}	& -	& \ensuremath{\begin{array}{c}\input{pstex/A-GR-combo-B.pstex_t} \end{array}}		& -	& \ensuremath{\begin{array}{c}\input{pstex/A-combo-GR-B.pstex_t} \end{array}},
	\end{array}
\ee
where
the arrow on the momentum derivative symbol indicates in which
sense the momentum derivative acts. This is unnecessary in
the parent diagram on the \lhs\ of~\eq{EP-dCTP-EP}, since
it is obvious that the momentum derivative has `pushed
forward' and so corresponds to a derivative
\wrt\ the momentum flowing \emph{into} the vertex,
from the structure $B$. However, once the vertex has been removed via
the effective propagator relation, this information is
lost, unless explicitly indicated.

Allowing the active gauge remainders in~\eq{EP-dCTP-EP} to act
(according to~\eq{WID-A}) leads us to define
\be
\label{eq:Combo}
	\ensuremath{\begin{array}{c}\begin{picture}(0,0)%
\includegraphics{pstex/Combo.pstex}%
\end{picture}%
\setlength{\unitlength}{3947sp}%
\begingroup\makeatletter\ifx\SetFigFont\undefined%
\gdef\SetFigFont#1#2#3#4#5{%
  \reset@font\fontsize{#1}{#2pt}%
  \fontfamily{#3}\fontseries{#4}\fontshape{#5}%
  \selectfont}%
\fi\endgroup%
\begin{picture}(313,629)(1277,1164)
\end{picture}%
 \end{array}} \equiv \ensuremath{\begin{array}{c}\begin{picture}(0,0)%
\includegraphics{pstex/EP-dGR.pstex}%
\end{picture}%
\setlength{\unitlength}{3947sp}%
\begingroup\makeatletter\ifx\SetFigFont\undefined%
\gdef\SetFigFont#1#2#3#4#5{%
  \reset@font\fontsize{#1}{#2pt}%
  \fontfamily{#3}\fontseries{#4}\fontshape{#5}%
  \selectfont}%
\fi\endgroup%
\begin{picture}(396,693)(1350,1119)
\end{picture}%
 \end{array}} -\hf \ensuremath{\begin{array}{c}\begin{picture}(0,0)%
\includegraphics{pstex/dPEP-GR.pstex}%
\end{picture}%
\setlength{\unitlength}{3947sp}%
\begingroup\makeatletter\ifx\SetFigFont\undefined%
\gdef\SetFigFont#1#2#3#4#5{%
  \reset@font\fontsize{#1}{#2pt}%
  \fontfamily{#3}\fontseries{#4}\fontshape{#5}%
  \selectfont}%
\fi\endgroup%
\begin{picture}(321,630)(893,1222)
\end{picture}%
 \end{array}},
\ee
which we will reqire shortly. Notice that the second term
of the \rhs\ contains a pseudo effective propagator
(differentiated \wrt\ to the momentum flowing into its
bottom end).

\section{The One-loop $\beta$ function}
\label{sec:beta1}

As an illustration of the formalism, we would like to
reproduce a standard result, namely the one-loop
$\beta$ function for massless QCD (we take the massless
case since, in our mass-dependent scheme, massive quarks will spoil
universality, rendering any comparison with other methods
of limited use). Fortunately, due to the developments
in~\cite{RG2005,mgiuc}, this calculation is extremely
easy. In~\cite{mgiuc} a diagrammatic 
expression has been
derived for the $n$-loop $\beta$ function in $\SU(N)$
Yang-Mills from which the universal answer (at least
at one and two loops) can be
directly extracted. The key to deriving this expression
is the effective propagator relation, which we have ensured
holds for QCD. Indeed, 
at the one loop level, the pure $\SU(N)$ Yang-Mills
expression
is exactly the same as in QCD, modulo the changes
to the Feynman rules, and it is given in \fig{beta1}.
(Beyond one loop, the expression of~\cite{mgiuc} is only
slightly modified.) We define $\Box_{\mu\nu}(p) \equiv p^2 \delta_{\mu \nu} - p_\mu p_\nu$
and take wiggly lines to denote physical gauge fields (with Lorentz indices
suppressed).
Note that in $D=4$ only the second, third and final diagrams
contribute, so the expression for $\beta_1$ is really very simple.
\bcf[h]
	\[
	\fl
	4 \beta_1 \Box_{\mu \nu} (p) + \Op{4}=
	-\frac{1}{2}
	\dec{
		\begin{array}{c}
			\begin{array}{ccccccc}
				\ensuremath{\begin{array}{c}\begin{picture}(0,0)%
\includegraphics{pstex/Beta1-LdL-A.pstex}%
\end{picture}%
\setlength{\unitlength}{3947sp}%
\begingroup\makeatletter\ifx\SetFigFont\undefined%
\gdef\SetFigFont#1#2#3#4#5{%
  \reset@font\fontsize{#1}{#2pt}%
  \fontfamily{#3}\fontseries{#4}\fontshape{#5}%
  \selectfont}%
\fi\endgroup%
\begin{picture}(512,510)(333,-63)
\put(557,140){\makebox(0,0)[lb]{\smash{\SetFigFont{11}{13.2}{\rmdefault}{\mddefault}{\updefault}{\color[rgb]{0,0,0}0}%
}}}
\end{picture}
 \end{array}}	& -	& \ensuremath{\begin{array}{c}\input{pstex/Beta1-LdL-B.pstex_t} \end{array}}	&+4	& \ensuremath{\begin{array}{c}\begin{picture}(0,0)%
\includegraphics{pstex/Beta1-LdL-C.pstex}%
\end{picture}%
\setlength{\unitlength}{3947sp}%
\begingroup\makeatletter\ifx\SetFigFont\undefined%
\gdef\SetFigFont#1#2#3#4#5{%
  \reset@font\fontsize{#1}{#2pt}%
  \fontfamily{#3}\fontseries{#4}\fontshape{#5}%
  \selectfont}%
\fi\endgroup%
\begin{picture}(378,371)(263,231)
\end{picture}
 \end{array}}	& -	&\ensuremath{\begin{array}{c}\input{pstex/Beta1-LdL-D.pstex_t} \end{array}}
			\end{array}
		\\[10ex]
			\begin{array}{cccc}
				+4	& \ensuremath{\begin{array}{c}\begin{picture}(0,0)%
\includegraphics{pstex/Beta1-LdL-E.pstex}%
\end{picture}%
\setlength{\unitlength}{3947sp}%
\begingroup\makeatletter\ifx\SetFigFont\undefined%
\gdef\SetFigFont#1#2#3#4#5{%
  \reset@font\fontsize{#1}{#2pt}%
  \fontfamily{#3}\fontseries{#4}\fontshape{#5}%
  \selectfont}%
\fi\endgroup%
\begin{picture}(512,835)(351,-443)
\put(572,-246){\makebox(0,0)[lb]{\smash{\SetFigFont{11}{13.2}{\rmdefault}{\mddefault}{\updefault}{\color[rgb]{0,0,0}0}%
}}}
\end{picture}
 \end{array}}	&-8	& \ensuremath{\begin{array}{c}\begin{picture}(0,0)%
\includegraphics{pstex/Beta1-LdL-Fc.pstex}%
\end{picture}%
\setlength{\unitlength}{3947sp}%
\begingroup\makeatletter\ifx\SetFigFont\undefined%
\gdef\SetFigFont#1#2#3#4#5{%
  \reset@font\fontsize{#1}{#2pt}%
  \fontfamily{#3}\fontseries{#4}\fontshape{#5}%
  \selectfont}%
\fi\endgroup%
\begin{picture}(288,1141)(608,-537)
\put(687,-241){\makebox(0,0)[lb]{\smash{{\SetFigFont{11}{13.2}{\rmdefault}{\mddefault}{\updefault}{\color[rgb]{0,0,0}0}%
}}}}
\end{picture}%
 \end{array}}
			\end{array}
		\end{array}
	}{\bullet}
	\]
\caption{Diagrammatic expression for $\beta_1$.}
\label{fig:beta1}
\ecf

From comparison with the Yang-Mills (YM) expression~\cite{mgierg1,mgierg2}
and the QED expression~\cite{giqed}, it is immediately clear
that, as in conventional approaches (though with the number of flavours 
set equal to twice
the number of colours),
\[
	\beta_1^{\mathrm{QCD}} = \beta_1^{\mathrm{YM}} + 2N\frac{\beta_1^{\mathrm{QED}}}{2}.
\]
Setting the quark masses to zero yields
\[
	\beta_1^{\mathrm{QCD}} = -\frac{N}{(4\pi)^2}\left(\frac{11}{3}-\frac{4}{3}\right). 
\]

\section{Conclusion}
\label{sec:conc}

We have constructed a manifestly gauge invariant
ERG for QCD and have used it to compute
the one-loop $\beta$ function for
$\SU(N)$ Yang-Mills coupled to $2N$ 
quarks. In the massless limit, we recovered 
the universal result.

The formalism is a direct extension of the
one developed for $\SU(N)$ Yang-Mills 
in~\cite{mgierg1}. The incorporation of
the quarks comprised three steps. First, the
quarks had to be added in a way that
respected the $\SU(N|N)$
regularization scheme. The symmetry associated
with the centre of this algebra in fact
prevented simply embedding the quarks
into the fundamental representation
of $\SU(N|N)$. Instead, we first embedded
sets of $N$
quarks into $N \times N$ matrices, whose rows
(columns)
were labelled by colour (flavour).
Each of these matrices was then embedded
into a separate supermatrix valued in
complexified $\U(N|N)$. In this way, we
were able to include multiples of $N$ quarks,
with each set of $N$ having degenerate masses.

The second step was to give each 
of the quarks independent masses, and this
required that we broke an unphysical, gauged $\SU(N)$
flavour symmetry which is carried by one
of the fields belonging to the $\SU(N|N)$
regularizing structure. To do this, we introduced
Higgs fields for each of the sets of $N$ quarks, whose
\vev\@s are essentially mass matrices.
The introduction of non-degenerate masses lifts
the restriction that the number of flavours must
be a multiple of the number of colours, since
we are at liberty to remove quarks from the
spectrum by tuning their masses to infinity. Clearly,
though, the construction is most efficient
when, suggestively, the number of flavours
is a multiple of the number of colours.

The third and final step 
was to adapt the ERG equation. This involved not
just including additional terms for the additional
fields but also defining the covariantization
of the ERG kernels appropriately. The key point
is that we require the number of independent
kernels to be equal to the number of independent,
propagating fields. Having given algebraic examples
of the covariantizations needed, we wrote down
the full QCD flow equation in its diagrammatic
form. Besides explicitly including the anomalous
dimensions of the renormalizing fields,
this expression has exactly the same form as that used
for pure $\SU(N)$ Yang-Mills. Indeed, this similarity
allowed us to directly write down a diagrammatic 
expression
for the one-loop $\beta$ function in QCD. By setting the
quark masses to zero we were able to recover
the standard, universal result.

The development of a manifestly gauge invariant ERG
for QCD is timely. In $\SU(N)$ Yang-Mills,
methods exist for computing the expectation values of
gauge invariant operators without fixing the 
gauge~\cite{evalues} and
this work can now be directly generalized to include
quarks.  We find
this particularly exciting
in view of the
fact that important progress is being made in understanding
the structure of non-perturbative contributions both to
these expectation values 
and the $\beta$ function~\cite{InPrep}.

\ack

OJR acknowledges financial support from PPARC.

\end{document}